\documentclass[11pt]{article}
\usepackage{latexsym,amsmath,amscd,amssymb, graphics}
\usepackage{amsmath,amssymb,mathrsfs,framed,esint,slashed,color}
\usepackage{amsthm}
\usepackage{enumerate}
\usepackage[colorlinks]{hyperref}
\usepackage{enumitem}
\usepackage[margin=1in]{geometry}
\usepackage{stmaryrd}
\usepackage{graphicx}
\usepackage[all]{xy}
\usepackage{pgfplots, pgfplotstable, booktabs, colortbl, array,multirow}

\definecolor{sand}{rgb}{0.76, 0.7, 0.5}
\definecolor{taupegray}{rgb}{1, 0, 0}
\usepackage[colorlinks]{hyperref}
\hypersetup{%
pdfstartview={FitH},
urlcolor=sand,
citecolor=blue,
linkcolor=taupegray,
}

\usepackage{framed}

\theoremstyle{plain}
\newtheorem{theorem}{Theorem}[section]

\theoremstyle{definition}

\newtheorem{remark}{Remark}[section]

\newcommand{\todo}[1]{\vspace{5 mm}\par \noindent
\framebox{\begin{minipage}[c]{0.95 \textwidth} \tt #1
\end{minipage}}\vspace{5 mm}\par}

\def\be{\begin{equation}}
\def\ee{\end{equation}}
\def\bea{\begin{eqnarray}}
\def\eea{\end{eqnarray}}
\def\ba{\begin{array}}
\def\ea{\end{array}}

\def\boldeta{\boldsymbol{\eta}}
\def\brho{\boldsymbol{\rho}}

\def\bx{{\mathbf {x} }}

\def\div{\mbox{div}\,}

\let\<\langle
\let \>\rangle

\newcommand{\rem}[1]{}

\newcommand{\de}{\delta}

\newcommand{\bvarphi}{\boldsymbol{\varphi}}

\newcommand{\bm}{\boldsymbol{m}}

\newcommand{\bu}{\boldsymbol{u}}

\newcommand{\bv}{\boldsymbol{v}}
\newcommand{\bX}{\boldsymbol{X}}

\newcommand{\bpsi}{\boldsymbol{\Psi}}

\newcommand{\bxi}{\boldsymbol{\xi}}

\newcommand{\bF}{\boldsymbol{F}}

\newcommand{\bk}{\mathbf{k}}
\newcommand{\bn}{\mathbf{n}}

\newcommand{\pp}[2]{\frac{\partial #1}{\partial #2}}

\newcommand{\dede}[2]{\frac{\delta #1}{\delta #2}}

\newcommand{\id}{\,\mathrm{Id}\,}

\newcommand{\tr}[1]{\mathrm{Tr}(#1)}

\begin{document}

\title{Geometric variational approach to the dynamics of porous media filled with incompressible fluid}

\author{Tagir Farkhutdinov\thanks{\noindent Department of mathematical and statistical sciences, University of Alberta, Edmonton, AB, T6G 2G1 Canada / ATCO SpaceLab, 5302 Forand St SW, Calgary, AB, T3E 8B4, Canada, \href{farkhutd@ualberta.ca}{farkhutd@ualberta.ca}}
\; and \; Fran\c{c}ois Gay-Balmaz\thanks{\noindent CNRS - LMD, Ecole Normale Sup\'erieure, 24 Rue Lhomond, 75005 Paris, \href{francois.gay-balmaz@lmd.ens.fr}{francois.gay-balmaz@lmd.ens.fr}}
\; and \; 
Vakhtang Putkaradze\thanks{\noindent Department of mathematical and statistical sciences, University of Alberta, Edmonton, AB, T6G 2G1 Canada / ATCO SpaceLab, 5302 Forand St SW, Calgary, AB, T3E 8B4, Canada, \href{putkarad@ualberta.ca}{putkarad@ualberta.ca}}
}

\date{}

\maketitle

\begin{abstract}
We derive the equations of motion for the dynamics of a porous media filled with an incompressible fluid. We use a variational approach with a Lagrangian written as the sum of terms representing the kinetic and potential energy of the elastic matrix, and the kinetic energy of the fluid, coupled through the constraint of incompressibility. As an illustration of the method, the equations of motion for both the elastic matrix and the fluid are derived in the spatial (Eulerian) frame. Such an approach is of relevance \emph{e.g.} for biological problems, such as sponges in water, where the elastic porous media is highly flexible and the motion of the fluid has a 'primary' role in the motion of the whole system.  We then analyze the linearized equations of motion describing the propagation of waves through the media. In particular, we derive the propagation of $S$-waves and $P$-waves in an isotropic media. We also analyze the stability criteria for the wave equations and show that they are equivalent to the physicality conditions of the elastic matrix. Finally, we show that the celebrated Biot's equations for waves in porous media are obtained for certain values of parameters in our models.
\end{abstract}

\tableofcontents

\section{Introduction}

The coupled dynamics of porous media filled with fluid, also known as \emph{poromechanics}, has been the subject of an active research for many decades. The foundational works in the area were driven by applications to soil dynamics and geophysics, whereas in the latter years the applications also included biomedical fields. The earlier developments were associated with the works of K. von Terzaghi \cite{Te1943} and M. Biot \cite{biot1941general,biot1955theory,biot1957elastic} in the consolidation of porous media, and subsequent works by M. Biot which derived the time-dependent equations of motion for poromechanics, based on certain assumptions on the media, and considered the wave propagation in both low and high wavenumber regime  \cite{biot1962mechanics,biot1962generalized,biot1963theory,biot1972theory}. There has been substantial amount of new work in the field of porous media, see \cite{joseph1982nonlinear,detournay1993fundamentals,dell1998micro,brovko2007continuum,carcione2010computational,grillo2014darcy} and subsequent mathematical analysis of the models \cite{showalter2000diffusion,bociu2016analysis,bastide2018penalization}. We refer the reader interested in the history of the field to the  review \cite{rajagopal2007hierarchy} for a more detailed exposition of the literature.

While Biot's equations, especially with respect to acoustic propagation in porous media, still remain highly influential today, subsequent investigations have revealed difficulties in the interpretation of various terms through the general principles of  mechanics, such as material objectivity, frequency-dependent permeability and changes of porosity in the model, as well as the need to describe large deformations of the model \cite{wilmanski2006few}. The above-cited paper then proceeds in outlining a detailed derivation for the modern approach to saturated porous media equations which does not have the limitations of the Biot's model. We shall also mention here two recent papers \cite{ChMo2010,ChMo2014} where the equations for saturated porous media were further developed based on the general thermodynamics principles of mechanics. 

By their very nature, variational methods involve fully nonlinear treatment of the inertial terms. The mainstream approach to the porous media has been to treat the dynamics as being friction-dominated by dropping the inertial terms from the equations. The equations we will derive here, without the viscous terms, will be of infinite-dimensional Hamiltonian type. On the other hand, the friction-dominated approach gives equations of motion that are of gradient flow type. The seminal book of Coussy \cite{coussy1995mechanics} contains a lot of background information and analysis. For more recent work, we will refer the reader to, for example, the studies of multi-component porous media flow  \cite{seguin2019multi}, 
as well as the gradient approach to the thermo-poro-visco-elastic processes \cite{both2019gradient}.

Fluid-filled elastic porous media, by its very nature, is a highly complex object involving both the individual dynamics of fluid and media, and a highly nontrivial interactions between them. The pores in the elastic matrix, and the fluid motion inside them, are micro-structured elements that contribute to the macro-structured dynamics. Thus, the porous media must include the interaction between the large scale dynamics and an accurate, and yet treatable, description of micro-structures. It has long been known in mechanics that variational principles are ideally suited to treat complex, multi-component systems. Variational methods proceed formally by describing the Lagrangian of the system on an appropriate configuration manifold, and proceeding with variations to obtain the equations of motion in a systematic way. The advantage of the variational methods is their consistency, as opposed to the theories based on balancing the conservation laws for a given point, or volume, of fluid. In a highly complex system like poromechanics, especially when written in the non-inertial Lagrangian frame associated with the matrix, writing out all the forces and torques to obtain correct equations is very difficult. In contrast, the equations of motion, as well as the conservation laws, come out of variational methods automatically without the need to find all the forces and torques involved. 
 Thus, porous media looks like an ideal application for applications of variational principles. Before we proceed further, however, we would like to give a verbatim quote of an inspiring sentence from the conclusion of \cite{wilmanski2006few}: 
\\
\emph{It seems to be also clear that it is a waste of effort to try to construct a true variational principle as the Biot model contains a nonequilibrium variable, the increment of fluid contents which rules out the existence of such a principle.}
\\
In spite of this difficulty, variational methods were actively applied to the field poromechanics. One of the earliest papers papers in the field was \cite{bedford1979variational} where the kinetic energy of expansion was incorporated into the Lagrangian to obtain the equations of motion. In that work, several Lagrange multipliers were introduced to enforce the continuity equation for both solid and fluid.
The works 
\cite{aulisa2007variational,aulisa2010geometric} use variational principles for explanation of the Darcy-Forchheimer law. Furthermore, 
\cite{lopatnikov2004macroscopic,lopatnikov2010poroelasticity} derive the equations of porous media using additional terms in the Lagrangian coming from the kinetic energy of the microscopic fluctuations. 
Of particular interest to us are the works on the Variational Macroscopic Theory of Porous Media (VMTPM) which was formulated in its present form in 
\cite{dell2000variational,sciarra2008variational,madeo2008variational,dell2009boundary,serpieri2011formulation,serpieri2015variationally,serpieri2016general,auffray2015analytical,serpieri2016variational,travascio2017analysis}, also summarized in a recent book \cite{serpieri2017variational}. In these works, the microscopic dynamics of capillary pores is modelled by a second grade material, where the internal energy of the fluid depends on both the deformation gradient of the elastic media, and the gradients of local fluid content.  The study of a pre-stressed system using variational principles and subsequent study of propagation of sound waves was undertaken in \cite{placidi2008variational}.

One of the main assumptions of the VMTPM is the dependence of the internal energy of the fluid on the quantity measuring the micro-strain of the fluid, or, alternatively, the fluid content or local density of fluid, including, in some works, the gradients of that quantity. This assumption is physically relevant for compressible fluid, but, in our view, for an incompressible fluid (which, undoubtedly, is a mathematical abstraction), such dependence is difficult to interpret. For example, for geophysical applications, fluids are usually considered compressible because of the large pressures involved. In contrast, for biological applications like the dynamics of highly porous sponges in water, the compressibility effects can be neglected. For a truly incompressible fluid, it is difficult to assign a physical meaning to the dependence of internal energy of the fluid on the parameters of the porous media. We refer the reader to the the classical Arnold's description of incompressible fluid \cite{arnold1966geometrie} as geodesic motion on the group of volume-preserving diffeomorphism in the three-dimensional space, in the absence of external forces. In that theory the Lagrangian is simply the kinetic energy, as the potential energy of the fluid is absent, and the fluid pressure enters the equations from the incompressibility condition. The main result of the present paper is to extend this geometric description to the motion of the fluid-filled porous media, for the case when the fluid inside the pores is incompressible, and, neglecting all thermal effects,  without considering the internal energy of fluid. 

Before we delve into detailed derivations, it is useful to have a discussion on the physics of what is commonly considered the saturated porous media. In most, if not all, previous works, the saturated porous media is a combined object consisting of an (elastic) dense matrix, and a network of small connected pores filled with fluid. The fluid encounters substantial resistance when moving through the pores due to viscosity and the no-slip condition on the boundary. In such a formulation, it is easier to consider the motion of the porous matrix to be 'primary', and the motion of the fluid to be computed with respect to the porous matrix itself. Because the motion of the elastic matrix is 'primary', the equations are written in the system of coordinates consistent with the description of the elastic media, which is the material frame associated with the media. In this paper, we take an alternative view where we choose the same coordinate system of the stationary observer (Eulerian frame) for the description of both the fluid and the elastic media. Such system is more frequently used in the classical fluid description, but is less common in the description of elastic media. 
Physically, our description is more relevant for the case of a porous media consisting of a dense network of elastic 'threads' positioned inside the fluid, which is a case that has not been considered before. In our formulation, we choose the Eulerian description for both the fluid and the elastic matrix. It is worth noting that the combined Eulerian description is also applicable to the regular  porous media with a 'dense' matrix, and is also well suited for the description of wave propagation in such media. Finally, we shall also point out that our theory can be reformulated and is applicable for the familiar choice of the Lagrangian material description with respect to the elastic porous matrix. These descriptions are completely equivalent from the mathematical point of view, and this is rigorously justified by using the process of Lagrangian reduction by symmetry in continuum mechanics \cite{GBMaRa12}.

\section{Equations of motion for porous media in spatial coordinates} 
\label{sec:3D_eqs}

In this Section we derive the equations of motion for a porous medium filled with an incompressible fluid by using a variational formulation deduced from Hamilton's principle. We will follow the description of both fluid and elastic matrix, individually, as outlined in the book by Marsden and Hughes \cite{marsden1994mathematical}, where the reader can find the background and fill in technical details of the description of each media.

\subsection{Definition of variables}

We shall remark that the preferred description for the motion of an elastic body is achieved through the Lagrangian coordinates of the media as being the independent variables, and balancing the forces in the spatial frame or the  frame attached to the media. On the other hand, the description of the fluid equation is traditionally done in the Eulerian (spatial) frame. The combined mixed fluid-material motion for porous media can thus be described in either frame. In order to connect with the earlier works by Biot and subsequent analysis of wave propagation in the porous media, we compute the equations of motion in spatial coordinates throughout the paper.

\paragraph{Configuration of the elastic body and the fluid.} Suppose that at $t=0$ the fluid and the elastic body occupy completely a given volume ${\mathcal B} \subset \mathbb{R}^3$. By default, we are working with a three-dimensional system, although the equations of motion reduce trivially to the two- and one-dimensional cases. The motion of the elastic body (indexed by $s$) and the fluid (indexed by $f$) is defined by two time dependent maps $\bpsi$ and $\bvarphi$ defined on $\mathcal{B}$ with values in $\mathbb{R}^3$, with variables denoted as 
$\bx=\bpsi(t,\bX_s)$ and $\bx=\bvarphi(t,\bX_f)$. We assume that there is no fusion of either fluid or elastic body particles, so the map $\bpsi$ and $\bvarphi$ are embeddings for all times $t$, defining uniquely the mappings $\bX_s=\bpsi^{-1}(t,\bx)$ and 
$\bX_f=\bvarphi^{-1}(t,\bx)$.  We also assume that the fluid cannot escape the porous medium
or create voids, so at all times $t$, the domains occupied by the fluid  ${\mathcal B}_{t,f}=\bvarphi(t,\mathcal{B})$ and the elastic body ${\mathcal B}_{t,s}=\bpsi(t,\mathcal{B})$ coincide: 
${\mathcal B}_{t,f}={\mathcal B}_{t,s}={\mathcal B}_{t}$. Finally, we shall assume for simplicity that the domain ${\mathcal B}_t$ does not change with time, and will simply call it ${\mathcal B}$, hence both $\bvarphi$ and $\bpsi$ are diffeomorphisms of $\mathcal{B}$ for all time $t$. An extension to the case of a moving boundary is possible, although it will require appropriate modifications in the variational principle.

\paragraph{Velocities of the elastic body and the fluid.} 
The fluid velocity $\bu_f$ and elastic solid velocity $\bu_s$, measured relative to the fixed coordinate system, \emph{i.e.}, in the Eulerian representation, are given as usual by 
\begin{equation} 
\bu_f(t,\bx)=\partial_t  \bvarphi \big(t, \bvarphi^{-1}(t,\bx)\big) \, , \quad  \bu_s(t,\bx)=\partial_t \bpsi \big(t, \bpsi^{-1}(t,\bx)\big) \, ,
\label{vel_def} 
\end{equation} 
for all $\bx \in \mathcal{B}$.  Note that since $\bvarphi$ and $\bpsi$ keep the boundary $\partial \mathcal{B}$ invariant, the vector fields $\bu_f$ and $\bu_s$ are tangent to the boundary, \emph{i.e.},
\begin{equation} \label{free_slip} 
\bu_f\cdot\bn=0\, , \quad  \bu_s\cdot\bn=0 \, ,
\end{equation}
where $\bn$ is the unit normal vector field to the boundary.  One can alternatively impose that $\bvarphi$ and $\bpsi$ (or only $\bpsi$) keeps the boundary $ \partial \mathcal{B} $ pointwise fixed. In this case, one gets no-slip boundary conditions
\begin{equation} \label{no_slip} 
\bu_f|_{ \partial \mathcal{B} }=0\, , \quad  \bu_s|_{ \partial \mathcal{B} }=0 \, , \quad \text{(or only $ \bu_s|_{ \partial \mathcal{B} }=0$)}.
\end{equation}

\paragraph{Elastic deformations of the dry media.} 
In order to incorporate the description of the elastic deformations of the media in the potential energy, we consider the deformation gradient of $\bpsi$ denoted
\begin{equation} 
\mathbb{F}(t,\bX_s)=\nabla\bpsi(t,\bX_s)\, . 
\label{F_def} 
\end{equation} 
In the spatial frame, we consider the Finger deformation tensor $b(t,\bx)$ defined by
\begin{equation} 
b(t,\bx)=\mathbb{F}\,\mathbb{F}^\mathsf{T}(t,\bX_s)   \, ,  
\label{b_def} 
\end{equation} 
where $\bx= \bpsi(t,\bX_s)$, see the paragraph below for the intrinsic geometric definition of $b$. In coordinates, we have
\[
\mathbb{F}^i_A= \frac{\partial\bpsi^i}{\partial X^A_s},\qquad b^{ij}=  \frac{\partial\bpsi^i}{\partial X^A_s}\frac{\partial\bpsi^j}{\partial X^A_s}
\]
with the summation over $A$ is assumed. 

In general the deformation of an elastic media \emph{without fluid} leads to $b \neq \id$ (the unit tensor). The potential energy $V$  of deformation of the dry media thus depends on $b$. However, in our case there is another part that leads to the elastic potential energy, namely, the microscopic deformations of the pores that we shall describe below.

\paragraph{Internal deformation of the pores and constraint.} 
Let us now consider the volume occupied by the fluid in a given spatial domain. We assume that the fluid fills the pores completely, so the volume occupied by the fluid in any given spatial domain is equal to the net volume of pores in that volume. Let us take the  infinitesimal Eulerian volume $\mbox{d}^3 \bx$ and define the pore volume fraction $g(t,\bx)$, so that the volume of fluid is given by $g(t,\bx)\mbox{d}^3 \bx$. There are two aspects to take into account to obtain the available volume to the fluid, namely, the local concentration of pores $c(t,\bx)$ and the infinitesimal pore volume $v(t,\bx)$. 

If, for example, the pores are ``frozen" in the material, they simply move as material moves. Then, the change of the local concentrations of pores $c(t,\bx)$ due to deformations is given by
\begin{equation}\label{change_c}
c(t,\bpsi(t,\bX_s))|{\rm det} \,\mathbb{F}(t,\bX_s)| = c_0(\bX_s),
\end{equation}
where $c_0(\bX_s)$ is the initial concentration of pores in the Lagrangian point $\bX_s$. Using the definition \eqref{b_def} of the Finger tensor $b$ gives ${\rm det} \,b(t,\bx) = |{\rm det} \,\mathbb{F}(t,\bX_s)|^2$, hence we can rewrite the previous relation as
\[
c(t,\bx)\sqrt{{\rm det} \,b(t,\bx) } = c_0(\bX_s).
\]
In the case of an initially uniform porous media, \emph{i.e.}, $c_0=$ const, this formula shows that the concentration $c(t,\bx)$ is a function of the value $b(t,\bx)$ of the Finger deformation tensor
\begin{equation}
c(b)=\frac{c_0}{\sqrt{{\rm det}b}} \, .
\label{c_b_particular_neq} 
\end{equation}
Note that from \eqref{change_c}, the concentration of pores satisfies
\[
\partial_t c+ \operatorname{div}(c\bu_s)=0\,.
\]

On the other hand, the pores themselves can expand and contract, which one can understand as modeling the pores through infinitesimally small elastic volumes filled with fluid. When the pores expand, they generate stress in the material; however, the stress averaged over any volume that is much larger than the size of the pores, is going to vanish. We thus introduce an additional dependence of the elastic part of the media on the infinitesimal volume denoted $\mathcal{V}(t,\bX_s)$ in the Lagrangian description. Its Eulerian version is $v(t,\bx)$ with
\begin{equation}\label{relation_V_v}
v(t,\bpsi(t,\bX_s))= \mathcal{V}(t,\bX_s)\,.
\end{equation}

These two considerations lead to the following constraint on the total volume of pores, which is more easily written in the spatial description:
\begin{equation} 
g(t,\bx) = c(b(t,\bx)) v(t,\bx) \, . 
\label{g_c_v_constraint} 
\end{equation} 

\paragraph{Conservation law for the fluid.} 
In what follows, we will consider an incompressible fluid, as that case has not been studied in the literature in sufficient details. The density of the fluid itself is denoted as $\rho_f^0=$ const. We can thus discuss the conservation of the volume of fluid rather than the mass. Let us now look at the volume of fluid $g(t,\bx) \mbox{d}^3 \bx$ from a different point of view. The fluid must fill all the available volume completely, and it must have come from the initial point  $\bX_f=\bvarphi^{-1}(t,\bx)$. If the initial volume fraction at that point was $g_0(\bX_f) \mbox{d}^3 \bX_f$, then at a point $t$ in time we have 
\begin{equation} 
g(t,\bx) = g_0 \big(\bvarphi^{-1}(t,\bx)\big)J_{\bvarphi^{-1}}(t,\bx)  \, , \quad 
J_{\bvarphi^{-1}} := {\rm det} \big(\nabla \bvarphi^{-1}\big) \, . 
\label{cons_law_fluid} 
\end{equation} 
Differentiating \eqref{cons_law_fluid}, we obtain the conservation law for $g(t,\bx)$ written as 
\begin{equation} 
\partial_t g + \mbox{div} ( g\,\mathbf{u}_f ) =0 \, . 
\label{g_cons} 
\end{equation} 
The mass of the fluid in the given volume is $\rho_f^0 g \mbox{d}^3 \bx$. 
Note that the incompressibility condition of the fluid \emph{does not} mean that $\operatorname{div}\bu_f=0$. That statement is only true for the case where no elastic matrix is present, \emph{i.e.}, for pure fluid. In the porous media case, a given spatial volume contains both fluid and elastic parts. The conservation of volume available to the fluid is thus given by \eqref{g_cons}.

\paragraph{Conservation law for the elastic body.} The mass density of the elastic body, denoted $\rho_s$, satisfies an equation analogous to \eqref{cons_law_fluid}, namely,
\begin{equation} 
\rho_s(t,\bx) = \rho_{s,0}\big(\bpsi^{-1}(t,\bx)\big)J_{\bpsi^{-1}}(t,\bx)  \, ,
\label{cons_law_elastic} 
\end{equation} 
where $\rho_{s,0}(\bX_s)$ is the mass density in the reference configuration. The corresponding differentiated form is
\begin{equation} 
\partial_t \rho_s + \mbox{div} ( \rho_s \mathbf{u}_s ) =0 \, . 
\label{rho_s_cons} 
\end{equation}

\paragraph{Intrinsic geometric formulation.} To understand the transport equation of the Finger deformation tensor, it is advantageous to reformulate geometrically its definition. We assume that the reference configuration $\mathcal{B}$ is endowed with a reference Riemannian metric $G$, locally denoted $G=G_{AB} dX_s^AdX_s^B$ and we consider its inverse $G^{-1}$. It is a symmetric two-contravariant tensor locally denoted $G^{-1} =G^{AB}\frac{\partial}{\partial X_s^A}\frac{\partial}{\partial X_s^B}$ with $G^{AB}G_{BC}=\delta^A_C$. Then, the Finger deformation tensor is the symmetric two-contravariant tensor obtained by pushing forward $G^{-1}$ by the elastic configuration $\bpsi$, namely
\begin{equation}\label{intrinsic_def_b}
b= \bpsi_* G^{-1}\,.
\end{equation}
For a domain in three-dimensional Euclidian space, the Riemannian metric is simply an identity, and is often not included in the considerations. However, the differential-geometric considerations here are important, \emph{e.g.} for evolution of porous shells, which we do not consider here. The geometric description presented here is explained in details in \cite{marsden1994mathematical}.
Using local coordinates, one notes that when $G$ is the Euclidean metric,  \eqref{intrinsic_def_b} reduces to \eqref{b_def}.
Using \eqref{intrinsic_def_b} and \eqref{vel_def}, we get the transport equation for $b$ as
\[
\partial_t b + \pounds_{\bu_s} b=0\,,
\]
where $\pounds_{\bu_s}$ denotes the Lie derivative of a two-contravariant tensor, given in coordinates by
\begin{equation}\label{Lie_der_b}
(\pounds_{\bu_s}b)^{ij}= \frac{\partial b^{ij}}{\partial x^k} u_s^k - b^{kj}\frac{\partial u_s^i }{\partial x^k} - b^{ik}\frac{\partial u_s^j }{\partial x^k}\,. 
\end{equation}

Let us now formulate \eqref{c_b_particular_neq} intrinsically \emph{i.e.}, without the use of the local coordinates. Given a Riemanian metric $\gamma$ on the spatial domain, the Jacobian $J_{\bpsi}$ of $\bpsi$ is defined by $\bpsi^*\mu_\gamma= J_{\bpsi} \mu_G$, where $\mu_\gamma=\sqrt{\det \gamma} \,{\rm d}^3\bx$ and $\mu_G= \sqrt{\det G} \,{\rm d}^3\bX_s$ are the Riemannian volume forms. From this, one expresses intrinsically the Jacobian of $\bpsi$ in terms of the Finger deformation tensor as
\[
J_{\bpsi} \circ \bpsi^{-1}= \frac{\mu_\gamma}{\mu_{b^{-1}}},
\]
where the Riemannian metric $b^{-1}$ is the inverse of $b$. 

Since equation \eqref{change_c} can be written intrinsically as $(c\circ \bpsi) J_{\bpsi}= c_0$, we get
\[
c= \frac{c_0\circ\bpsi^{-1}}{J_{\bpsi}\circ\bpsi^{-1}}= (c_0\circ\bpsi^{-1})\frac{\mu_{b^{-1}}}{\mu_\gamma}.
\]
If $c_0=$ const, we get the expression
\[
c(b)= c_0\frac{\mu_{b^{-1}}}{\mu_\gamma}
\]
which is the intrinsic version of \eqref{c_b_particular_neq}.

\paragraph{Summary of the variables in the Lagrangian and Eulerian descriptions.} From the discussion above, the independent variables in the Lagrangian descriptions are the two embeddings and the infinitesimal volume, \emph{i.e.},
\begin{equation}\label{Lagrangian_variables}
\bpsi(t,\bX_s),\qquad \bvarphi(t,\bX_f), \qquad \mathcal{V}(t,\bX_s).
\end{equation}
In the Eulerian description the variables are
\begin{equation}\label{Eulerian_variables}
\bu_f(t,\bx),\qquad \bu_s(t,\bx),\qquad v(t,\bx),\qquad g(t,\bx), \qquad\rho_s(t,\bx),\qquad b(t,\bx)\,,
\end{equation}
defined from the Lagrangian variables in \eqref{vel_def}, \eqref{relation_V_v}, \eqref{cons_law_fluid}, \eqref{cons_law_elastic}, \eqref{intrinsic_def_b}, respectively.

\subsection{Lagrangian and variational principle in spatial variables}

\paragraph{Lagrangian.} For classical elastic bodies, the potential energy in the spatial description depends on the Finger deformation tensor $b$, \emph{i.e.}, $V=V(b)$. If the pores are present and their volumes change due to their expansion or contraction, the potential energy changes even when there are no net deformation of the porous media. Thus, the potential energy of elastic porous material must depend on both variables $b$ and $v$, and we write $V=V(b,v)$.

The Lagrangian of the porous medium is the sum of the kinetic energies of the fluid and elastic body minus the potential energy of the elastic deformations: 
\begin{equation} 
\ell(\bu_f,\bu_s, \rho_s,b,g,v)= \int_{\mathcal{B}}
\left[\frac{1}{2}  \rho_f  g |\bu_f|^2 + \frac{1}{2} \rho_s |\bu_s|^2 -V(b,v)\right]{\rm d}^3\bx \,. 
\label{Lagr_def} 
\end{equation} 
Note that the expression \eqref{Lagr_def} explicitly separates the contribution from the fluid and the elastic body in simple physically understandable terms. The interaction between the fluid and the media comes from the critical action principle involving the incompressibility of the fluid. We shall derive the equations of motion for an arbitrary  (sufficiently smooth) expression for $\ell(\bu_s,\bu_f, \rho_s,b,g,v)$, and will use 
 the physical Lagrangian \eqref{Lagr_def} for all computations in the paper.

\paragraph{Variational principle and incompressibility constraint.} Condition \eqref{g_c_v_constraint} represents a scalar constraint for every point of an infinite-dimensional system. Formally, such constraint can be treated in terms of Lagrange multipliers. The application of the method of Lagrange multipliers for an infinite-dimensional system is quite challenging, see recent review papers \cite{dell2018lagrange,bersani2019lagrange}. In terms of classical fluid flow, in the framework of Euler equations, the variational theory introducing incompressibility constraint has been developed by V. I. Arnold \cite{arnold1966geometrie}, with the Lagrange multiplier for incompressibility related to the physical pressure in the fluid. We will follow in the footsteps of Arnold's method and introduce a Lagrange multiplier for the incompressibility condition \eqref{g_c_v_constraint}. By analogy with Arnold, we will also treat this Lagrange multiplier as related to pressure, as it has the same dimensions, and denote it $p$.  Since \eqref{g_c_v_constraint} refers to the fluid content, the Lagrange multiplier $p$ relates to the fluid pressure. This will be further justified by the equations of motion \eqref{eq_gen} below, connecting pressure with the derivatives of the potential energy with respect of the pores' volume.   Note that $p$ may be different from the actual physical pressure in the fluid depending on the implementation of the model. From the Lagrangian \eqref{Lagr_def} and the constraint  \eqref{g_c_v_constraint}, we define the action functional in the Eulerian description as
\begin{equation} 
S= \int_0^T\left[ \ell(\bu_f,\bu_s, \rho_s,b,g,v) -  \int_{{\cal B} } p\big( g- c(b) v\big) \mbox{d}^3 \bx\right] \mbox{d} t \,. 
\label{action_p}
\end{equation}

The equations of motion are obtained by computing the critical points of $S$ with respect to constrained variations of the Eulerian variables induced by free variations of the Lagrangian variables. Indeed, it is in the Lagrangian description that the variational principle is justified, as being given by the Hamilton principle with constraint. One also notes that the constraint \eqref{g_c_v_constraint} is holonomic when expressed in terms of the Lagrangian variables \eqref{Lagrangian_variables} via the relations \eqref{relation_V_v}, \eqref{cons_law_fluid}, \eqref{intrinsic_def_b}. This justifies that this constraint can be incorporated via the introduction of a Lagrange multiplier.
The constrained variations of the Eulerian variables induced by the free variations $\delta\bpsi$, $\delta\bvarphi$ vanishing at $t=0,T$  are computed by using the relations \eqref{vel_def}, \eqref{cons_law_fluid}, \eqref{cons_law_elastic}, \eqref{intrinsic_def_b}. This yields
\begin{equation} 
\begin{aligned}
\de \bu_f &= \partial_t \boldeta_{f} + \bu_f  \cdot \nabla\boldeta_f - 
\boldeta_f \cdot  \nabla \bu_f \\ 
\de \bu_s &= \partial_t \boldeta_{s} + \bu_s  \cdot \nabla\boldeta_s - 
\boldeta_s \cdot  \nabla\bu_s \\
\delta g&= -  \operatorname{div}(g\boldeta_f) \\ 
\delta\rho_s&= - \operatorname{div}(\rho_s\boldeta_s) \\ 
\delta b &= - \pounds_{\boldeta_s}b \, ,
\label{g_rhos_b_var}
\end{aligned} 
\end{equation} 
where  $\boldeta_{f}$ and $\boldeta_{s}$ are defined
\begin{equation} 
\boldeta_{f}=\de \bvarphi \circ \bvarphi^{-1} \, , \quad 
\boldeta_{s}=\de \bpsi \circ \bpsi^{-1} 
\label{eta_def} 
\end{equation} 
and the variations $\delta v$ and $\delta p$ are arbitrary.  In the case of the boundary conditions \eqref{free_slip} it follows from \eqref{eta_def} that $\boldeta_{f}$ and $\boldeta_{s}$ are  arbitrary time dependent vector fields vanishing at $t=0,T$ and tangent to the boundary $\partial\mathcal{B}$:
\begin{equation} 
\label{non-permeable} 
\boldeta_s\cdot\bn=0\, , \quad  \boldeta_f\cdot\bn=0 \, . 
\end{equation}
In the case of no-slip boundary conditions \eqref{no_slip}, we have
\begin{equation} 
\label{no_slip_var} 
\boldeta_f|_{ \partial \mathcal{B} }=0\, , \quad  \boldeta_s|_{ \partial \mathcal{B} }=0 \, , \quad \text{(or only $\boldeta_s|_{ \partial \mathcal{B} }=0$)}.
\end{equation}

\paragraph{Incorporation of external and friction forces.} Frictions forces, or any other forces, acting on the fluid $\bF_f$ and the media $\bF_s$ can be incorporated into the variational formulation by using the Lagrange-d'Alembert principle for external forces.  This principle reads
\begin{equation} 
\de S+ \int_{{\cal B}}\left(  \bF_f \cdot \boldeta_f + \bF_s \cdot \boldeta_s \right)  \mbox{d}^3\bx \, \mbox{d} t =0 \, , \quad  
\label{Crit_action} 
\end{equation} 
where $S$ is defined in \eqref{action_p} and the variations are given by \eqref{g_rhos_b_var}. 
Such friction forces are usually postulated from general physical considerations. If these forces are due exclusively to friction, the forces acting on the fluid and media at any given point must be equal and opposite, \emph{i.e.} $\bF_f=-\bF_s$, in the Eulerian treatment we consider here. For example, for porous media it is common to posit the friction law
\begin{equation} 
\bF_f = - \bF_s= \mathbb{K} (\bu_s - \bu_f)\,,
\label{Darcy_law} 
\end{equation} 
with $\mathbb{K}$ being a positive definite matrix potentially dependent on material parameters and variables representing the media. In particular, the matrix $\mathbb{K}$ depends on the local porosity, composition of the porous media, deformation and other variables. The general functional form of dependence of $\mathbb{K}$ on the variables should be of the form $\mathbb{K}=\mathbb{K}(b,g)$. 
For example, when  deformations of porous media are neglected, \emph{i.e.}, assuming and isotropic and a non-moving porous matrix with $b={\rm Id}$, Kozeny-Carman equation is often used, which in our notation is  written in the form  $\mathbb{K} = \kappa g^3/(1-g)^2$, with $\kappa$ being a constant, see \cite{costa2006permeability} for discussion.
In general, the derivation of the dependence of tensor $\mathbb{K}$ on variables $g$ and $b$ from the first principles is difficult, and should presumably be obtained from experimental observations. In general, the anisotropy of $\mathbb{K}$ is related to the geometry of the pores.  The shape of the pores and their distribution in space will dictate the numerical values of $\mathbb{K}$ for each given point in space, and the deformation of the pores' geometry will determine the functional dependence $\mathbb{K}=\mathbb{K}(b,g)$. For the purpose of this paper, we will implicitly assume the dependence on flow variables without specifying them explicitly in the formulas. For computations in  Section~\ref{sec:lin_stab} dedicated to the description of propagation of linear disturbances about the steady state, such dependence of $\mathbb{K}$ on variables is not important. If there are other external forces acting on the system, then, in general, $\bF_f + \bF_s \neq 0$. This situation can happen, for example, if either the fluid or the media is either electrostatically charged or laden with magnetic particles, and is subjected to the electric or magnetic field. The equations that we derive in the general setting are valid for arbitrary external forces $\bF_f$ and $\bF_s$. For explicit computations, we assume the expression \eqref{Darcy_law}. 

\subsection{Equation of motion}

\paragraph{General form of the equations of motion.} In order to derive the equations of motion, we take the variations in the Lagrange-d'Alembert principle \eqref{Crit_action} as 
\begin{equation} 
\label{Crit_action_explicit} 
\begin{aligned} 
&\de S+ \int_{{\cal B}} \left( \bF_f \cdot \boldeta_f + \bF_s \cdot \boldeta_s \right)  \mbox{d}^3\bx\,  \mbox{d} t 
\\ 
&= \int_{{\cal B}} \left[ \dede{\ell}{\bu_f} \cdot \delta  \bu_f 
+  \dede{\ell}{\bu_s} \cdot \delta \bu_s + \dede{\ell}{\rho_s} \de \rho_s + \left( \dede{\ell}{b}+ pv \pp{c}{b} \right) :\de b  \right. 
\\ 
&  \textcolor{white}{\frac{1}{2} } \qquad\qquad + \left( \dede{\ell}{g} - p\right) \de g  
+  \left( \dede{\ell}{v} + p c(b)\right) \de v+  \big(g-c(b) v\big)  \de p \\ 
 & \left.  \textcolor{white}{\frac{1}{2} } \qquad\qquad +\bF_f \cdot \boldeta_f + \bF_s \cdot \boldeta_s \right]  \mbox{d}^3 \bx \,  \mbox{d} t=0 \,.
\end{aligned} 
\end{equation} 
The symbol $``:"$ denotes the contraction of tensors on both indices. Substituting the expressions for variations \eqref{g_rhos_b_var}, integrating by parts to isolate the quantities $\boldeta_f$ and $\boldeta_s$, and dropping the boundary terms leads to the expressions for the balance of the linear momentum for the fluid and porous medium, respectively, written in the Eulerian frame. This calculation is tedious yet straightforward for most terms and we omit it here. The main difficulty is the calculation of the terms related to the evolution of the tensor $b$, which we now show in some details. 

Denoting by $\Pi$ the 2-covariant symmetric tensor field $\dede{\ell}{b}+ pv \pp{c}{b} $, we compute the fourth term on the right hand side of \eqref{Crit_action_explicit} by using \eqref{Lie_der_b} as follows: 
\begin{equation} 
\label{diamond_calc} 
\begin{aligned} 
\!\!\!\!\int_ \mathcal{B} \Pi : \delta b&=-\int_ \mathcal{B} (\Pi : {\pounds}_{\boldeta} b) {\rm d} ^3 \bx\\
&=  -\int_{\cal B} \Pi_{ij} \left( \frac{\partial b^{ij}}{\partial x^k} \eta^k - b^{kj}\frac{\partial \eta^i }{\partial x^k} - b^{ik}\frac{\partial \eta^j }{\partial x^k}\right) \mbox{d}^3\bx \\ 
&= - \int_{\cal B} 
\left( \Pi_{ij}  \frac{\partial b^{ij}}{\partial x^k} \eta^k+ \eta^i \pp{}{x^k} \left( \Pi_{ij} b^{kj} \right) 
+  \eta^j \pp{}{x^k} \left( \Pi_{ij} b^{ik} \right) \right)\mbox{d}^3 \bx \\
&\qquad + \int_ \mathcal{B}  \frac{\partial }{\partial x ^k  } \left(  \Pi _{ij} b^{kj}  \eta ^i + \Pi _{ij} b^{ik} \eta ^j \right) {\rm d}^3 \bx \\
&= -\int_{\cal B}
\left( \Pi_{ij} \pp{b^{ij}}{x^k}+\pp{}{x^i} \left( \Pi_{k j}b^{ij} \right) + \pp{}{x^j} \left(  \Pi_{i k}b^{ij}  \right) 
\right) \eta^k \mbox{d}^3 \bx \\
&\qquad + 2 \int_{ \partial \mathcal{B} }\Pi _{ij} b^{kj} \eta ^i n_k {\rm d} s\\
&= -\int_{\cal B} 
\left( \Pi_{ij} \pp{b^{ij}}{x^k} + 2 \pp{}{x^i} \left( \Pi_{k j}b^{ij} \right) \right) \eta^k \mbox{d}^3 \bx + 2 \int_{ \partial \mathcal{B} }(\Pi _{ij} b^{kj} n_k) \eta ^i  {\rm d} s\,,
\end{aligned} 
\end{equation} 
where in three-dimensional case, $n_i$ are the components of the normal vector $\bn$ \footnote{ For a general metric $G$, the rigorous statement is that $n_i$ is the one form associated to the normal vector field $\bn$ via the Riemannian metric.}.
For compactness of notation, we denote the one-form appearing in the first term above with the \emph{diamond} operator 
\begin{equation}\label{diamond_coord}
(\Pi\diamond b)_k= - \Pi_{ij} \pp{b^{ij}}{x^k} - 2 \pp{}{x^i} \left( \Pi_{k j}b^{ij} \right)
\end{equation}
whose coordinate-free form reads 
\begin{equation}  
 \Pi \diamond b = - \Pi : \nabla b - 2 {\rm div} \left( \Pi \cdot b \right) \,.
\label{diamond_coord_free} 
\end{equation} 
The result of \eqref{diamond_calc} thus reads
\begin{equation}\label{boundary_b} 
\int_ \mathcal{B} \Pi : \delta b= \int _ \mathcal{B} ( \Pi \diamond b) \cdot \boldsymbol{\eta} \,{\rm d} ^3 \bx + 2 \int_{ \partial \mathcal{B} } [ (\Pi \cdot  b ) \cdot \mathbf{n} ] \cdot \boldsymbol{\eta} \,  {\rm d} s.
\end{equation} 

The equations of motion also naturally involve the expression of the Lie derivative of a momentum density, whose global and local expressions are
\begin{equation}\label{Lie_der_momentum}
\begin{aligned} 
\pounds_{\bu}\bm &= \bu \cdot \nabla \bm + \nabla\bu ^\mathsf{T} \cdot \bm + \bm \operatorname{div}\bu\\
(\pounds_{\bu}\bm)_i&= \partial_j m_i u^j + m_j \partial_i u^j+ m_i \partial_j u^j\,.
\end{aligned} 
\end{equation}

With these notations, the Lagrange-d'Alembert principle \eqref{Crit_action_explicit} yields the system of equations
\begin{equation} 
\label{eq_gen} 
\left\{
\begin{array}{l}
\displaystyle\vspace{0.2cm}\partial_t\frac{\delta \ell}{\delta \bu_f}+ \pounds_{\bu_f} \frac{\delta \ell}{\delta \bu_f} = g \nabla \left( \frac{\delta  {\ell}}{\delta g}-  p\right)+\bF_f\\
\displaystyle\vspace{0.2cm}\partial_t\frac{\delta\ell}{\delta \bu_s}+ \pounds_{\bu_s} \frac{\delta\ell}{\delta \bu_s} =  \rho_s\nabla \frac{\delta\ell}{\delta \rho_s} + \left(\frac{\delta\ell}{\delta b}+ p v\frac{\partial c}{\partial b}\right)\diamond b+ \bF_s\\
\displaystyle\vspace{0.2cm} \frac{\delta\ell}{\delta v}= -  pc(b)\,,\qquad g= c(b)v\\
\vspace{0.2cm}\partial_tg + \operatorname{div}(g\bu_f)=0\,,\qquad\partial_t\rho_s+\operatorname{div}(\rho_s\bu_s)=0\,,\qquad \partial_tb+ \pounds_{\bu_s}b=0\,.
\end{array}\right.
\end{equation}
When the boundary conditions \eqref{no_slip} are used, no additional boundary condition arise from the variational principle. In the case of the free slip boundary condition \eqref{free_slip}, the variational principle yields the condition
\begin{equation}\label{BC_general} 
[\sigma _p \cdot \mathbf{n} ] \cdot \boldsymbol{\eta} =0,\quad \text{for all $ \boldsymbol{\eta} $ parallel to $ \partial \mathcal{B} $},
\end{equation} 
where
\begin{equation}\label{def_sigma_p} 
\sigma _p := -2  \left( \frac{\delta\ell}{\delta b}+ p v\frac{\partial c}{\partial b} \right)  \cdot b.
\end{equation}
This is shown by using \eqref{boundary_b}. Physically, the condition \eqref{BC_general} states that the force  $ \mathbf{t}=\boldsymbol{\sigma} \cdot \mathbf{n} $ exerted at the boundary must be normal to the boundary (free slip).

The first equation arises from the term proportional to $\boldeta_f$  in the application of the Lagrange-d'Alembert principle. The second condition and the boundary condition \eqref{BC_general} arise from the term proportional to $\boldeta_s$ and via the use of \eqref{boundary_b}. The third and fourth equations arise from the variations $\delta v$ and $\delta p$.
The last three equations follow from the definitions \eqref{cons_law_fluid}, \eqref{cons_law_elastic}, \eqref{intrinsic_def_b}, respectively.
In the derivation of \eqref{eq_gen}, we have used the fact that on the boundary $\partial\mathcal{B}$, $\boldeta_s$ and $\boldeta_f$ satisfy the boundary condition \eqref{non-permeable}.

\begin{remark}[Discussion of the form of the Lagrangian] 
{\rm 
Equations \eqref{eq_gen} allow for an arbitrary form of the dependence of the Lagrangian on the variables. The derivatives of the Lagrangian with respect to the variables entering \eqref{eq_gen}  should be considered to be variational derivatives. For example, if the integrand of the Lagrangian depends on both $\rho_s$ and its spatial derivatives $\nabla \rho_s$,  \emph{e.g.} 
\[ 
\ell = \int_{\mathcal{B}} \ell_0 ( \rho_s, \nabla \rho_s, \bu_s, \ldots) \mbox{d} \bx
\] 
then 
\[ 
\dede{\ell}{\rho_s}=\pp{\ell_0}{\rho_s}-\div \pp{\ell_0}{\nabla \rho_s} \, , 
\] 
and similarly with other variables such as $\bu_s$, $\brho_f$, $v$ \emph{etc}. 
Thus, equations \eqref{eq_gen} are capable of incorporating very general physical models of the porous media. However, it is important to note that in our model, we do not assume that the energy of the fluid depends on any kind of strain  measure of the solid or the fluid. The pressure $p$ in \eqref{eq_gen} is obtained purely from the action principle with the action \eqref{action_p}. In that sense, our paper follows the framework of fluid description due to Arnold \cite{arnold1966geometrie}. 
}
\end{remark} 

\paragraph{Specific form of the equations.} We now use the Lagrangian function $\ell$ defined in \eqref{Lagr_def} and compute the derivatives 
\begin{equation} 
\left\{ 
\label{explicit_derivs} 
\begin{aligned}
& \frac{\delta \ell}{\delta \bu_f}= \rho_f  g\bu_f 
\, , \qquad 
\frac{\delta \ell}{\delta \bu_s}= \rho_s\bu_s 
\, , \qquad 
\frac{\delta \ell}{\delta \rho_s}= \frac{1}{2}|\bu_s|^2\,, \\
&
\frac{\delta \ell}{\delta b}=-\frac{\partial V}{\partial b} 
\, , \qquad 
\frac{\delta \ell}{\delta g}=\frac{1}{2}\rho_f|\bu_f|^2 
\, , \qquad 
\frac{\delta \ell}{\delta v}=- \frac{\partial V}{\partial v} \, . 
\end{aligned}
\right. 
\end{equation} 
For the Lagrangian in \eqref{Lagr_def}, using \eqref{diamond_coord_free} and the third and fourth equations in \eqref{eq_gen}, the diamond term in \eqref{eq_gen} simplifies as
\begin{align*}
\left( - \frac{\partial V}{\partial b} + p v \frac{\partial c}{\partial b}\right)\diamond b&=- \left(  p v \frac{\partial c}{\partial b} - \frac{\partial V}{\partial b}\right):\nabla b - 2 \operatorname{div}\left[\left(  p v \frac{\partial c}{\partial b} - \frac{\partial V}{\partial b}\right)\cdot b\right]\\
&=  g\nabla p + \nabla \left( V - \frac{\partial V}{\partial v}v\right) - 2 \operatorname{div}\left[ \left(  p v \frac{\partial c}{\partial b} - \frac{\partial V}{\partial b}\right)\cdot b\right].
\end{align*}
Then, the equations of motions \eqref{eq_gen} become 
\begin{equation}
\label{expressions_explicit} 
\hspace{-3mm} 
\left\{
\begin{array}{l}
\displaystyle\vspace{0.2cm}\rho_f(\partial_t \bu_f+ \bu_f\cdot\nabla \bu_f )  = - \nabla  p + \frac{1}{g} \bF_f\\
\displaystyle\vspace{0.2cm}\rho_s (\partial_t \bu_s \!+\! \bu_s\cdot\nabla \bu_s) =g\nabla p \!+\! \nabla \left( V\! - \!\frac{\partial V}{\partial v}v\right) - 2 \operatorname{div} \left[\left(  p v \frac{\partial c}{\partial b}\! -\! \frac{\partial V}{\partial b}\right)\cdot b\right]\!+\! \bF_s\\
\displaystyle\vspace{0.2cm} \frac{\partial V}{\partial v}=  pc(b),\qquad g= c(b)v\\
\vspace{0.2cm}\partial_tg + \operatorname{div}(g\bu_f)=0,\qquad \partial_t\rho_s+\operatorname{div}(\rho_s\bu_s)=0,\qquad \partial_tb+ \pounds_{\bu_s}b=0.
\end{array}\right.
\end{equation}
together with the boundary condition \eqref{BC_general} in which the stress tensor $ \sigma _p $ in \eqref{def_sigma_p} reads 
\begin{equation} 
\label{sigma_p} 
\sigma_p= -2 \left(  p v \frac{\partial c}{\partial b}\! -\! \frac{\partial V}{\partial b}\right)\cdot b\,, \qquad (\sigma_p)_k^i= -2 \left(  p v \frac{\partial c}{\partial b^{kj}}\! -\! \frac{\partial V}{\partial b^{kj}}\right)b^{ij}\,.
\end{equation}

The divergence term in the media momentum equation (second equation above) is the analogue of the divergence of the stress tensor for an ordinary elastic media: This term, however, contains the contribution from both the potential energy and the fluid pressure. 

These equations define the coupled motion of an incompressible fluid and porous media. We are not aware of these equations having been derived before. 

\begin{remark}[Equations of motion with external equilibrium pressure]
{\rm If the media is subjected to a  uniform external pressure $p_0$, then the equations of motion are derived by changing the Lagrangian to $\ell_p \rightarrow \ell + (p-p_0) (g-c(b) v)$. In that case, equations \eqref{eq_gen}, and, similarly, \eqref{expressions_explicit} are altered by simply substituting $p-p_0$ instead of $p$. In what follows, we shall put $p_0=0$. 
}
\end{remark}

\subsection{Energy dissipation}
We are now going to proceed to prove that our model yields strict  dissipation of mechanical energy in the presence of friction forces. This is important in order to demonstrate that our derivation is physically consistent. Fortunately, variational methods are guaranteed to provide energy conservation for the absence of friction, and when the friction forces are introduced correctly, also guaranteed to provide energy dissipation.
Let us consider the energy density associated to the Lagrangian $\ell$ given by
\begin{equation}
e=\bu_f \cdot \dede{\ell}{\bu_f} + 
\bu_s \cdot \dede{\ell}{\bu_s}
+ 
\dot v \dede{\ell}{\dot v} - \mathcal{L} \, ,
\label{energy_density} 
\end{equation}
where $\mathcal{L}$ denotes the integrand of $\ell$. 
Note that in our case $\ell$ does not depend on $\dot v$ hence the third term vanishes.
For the general system \eqref{eq_gen}, and its explicit form \eqref{expressions_explicit}, to be physically consistent, we need to prove that in the absence of  forces $\bF_s$ and $\bF_f$, the total energy $E= \int_{\cal B} e\, \mbox{d}^3 \bx$ is conserved. When these forces are caused by friction,  we must necessarily have  
$\dot E \leq 0$. 

We begin by noticing the formula 
\begin{equation}\label{formula_energy}
\bu \cdot {\pounds}_{\bu} \bm= 
\bu \cdot \left( \bu \cdot \nabla \bm+ \nabla \bu^\mathsf{T}\cdot\bm+ 
\bm\operatorname{div} \bu \right)= {\rm div} \big( \bu \,(\bm\cdot \bu)  \big) \, ,
\end{equation}
which easily follows from its coordinates expression in \eqref{Lie_der_momentum}.
Then, using equation \eqref{formula_energy} and system \eqref{eq_gen}, we compute
\begin{equation}\label{computation_energy_balance}
    \begin{aligned}
    \partial_t e &= \bu_f \cdot \pp{}{t} \dede{\ell}{\bu_f} 
    + 
    \bu_s \cdot \pp{}{t} \dede{\ell}{\bu_s} 
    - \dede{\ell}{\rho_s}\partial_t \rho_s  - \dede{\ell}{b}: \partial_t b  -\dede{\ell}{g}  \partial_t g  -  \dede{\ell}{v}\partial_t v
    \\ 
    &= - {\rm div} \left[ 
    \bu_f \left( \bu_f \cdot \dede{\ell}{\bu_f}\right)
    +
    \bu_s \left( \bu_s \cdot \dede{\ell}{\bu_s} \right) 
    - \left( \dede{\ell}{g}  - p \right) g \bu_f \right.\\
    &\hspace{5cm}\left.- \dede{\ell}{\rho_s}   \rho_s \bu_s + 2 \bu_s\cdot \left( \frac{\delta \ell}{\delta b} + pv \frac{\partial c}{\partial b} \right) \cdot b 
    \right] 
    \\
    &\qquad+\left( \dede{\ell}{g}-p \right) 
    \partial_t g
    + 
    \dede{\ell}{\rho_s} \partial_t \rho_s 
    + 
    \left(\dede{\ell}{b}+ pv \frac{\partial c}{\partial b}\right) \partial_t  b  \\
    &\qquad- \dede{\ell}{\rho_s} \partial_t \rho_s - \dede{\ell}{b}:\partial_t b 
    - \dede{\ell}{g} \partial_t g - \dede{\ell}{v} \partial_t v  + \bu_s \cdot \bF_s 
    + \bu_f \cdot \bF_f  
    \\ 
    &= - {\rm div} \boldsymbol{J}-p \partial_t g 
    + pv \frac{\partial c}{\partial b}:\partial_t  b - \dede{\ell}{v} \partial_t v+ \bu_s \cdot \bF_s 
    + \bu_f \cdot \bF_f\,,
    \end{aligned}
\end{equation}
where we denoted by $\boldsymbol{J}$ the vector field in the brackets inside the div operator. The last term in these brackets has the local expression
\[
\left(2 \bu_s\cdot \left( \frac{\delta \ell}{\delta b} + pv \frac{\partial c}{\partial b}\right)\cdot b \right)^k= 2 \bu_s^i \left( \frac{\delta \ell}{\delta b_{ij}} + pv \frac{\partial c}{\partial b_{ij}}\right) b ^{jk} = - \sigma _p \cdot \mathbf{u} _s\,.
\]
The sum of the second, third, and fourth terms in last line of \eqref{computation_energy_balance} cancel thanks to the third and fourth equations in \eqref{eq_gen}.
We thus get the energy balance
\[
\partial_t e+ \operatorname{div} \boldsymbol{J} = \bu_s \cdot \bF_s 
    + \bu_f \cdot \bF_f\,.
\]
Thus, the balance of total energy is
\begin{equation}
\label{energy_total}
\dot E= \int_\mathcal{B}  \left( \bu_s \cdot \bF_s + \bu_f \cdot \bF_f \right)  \mbox{d}^3\bx - \int_{\partial\mathcal{B}} \boldsymbol{J}\cdot \bn \,{\rm d}s\, . 
\end{equation}
 From the boundary conditions \eqref{free_slip} and \eqref{BC_general} we have $ \mathbf{u}_s \cdot \mathbf{n} =0$, $ \mathbf{u} _ f \cdot \mathbf{n} =0$, and $ [\sigma _p \cdot \mathbf{n} ] \cdot \mathbf{u}_s =0$ on the boundary $ \partial \mathcal{B} $, so that $\mathbf{J}\cdot \mathbf{n} =\mathbf{0}$ at the boundary. In the case of the boundary conditions \eqref{no_slip}, we have $ \mathbf{J}|_{ \partial \mathcal{B} }=0$.
In the absence of external forces, when $\bF_f$ and $\bF_s$ are caused exclusively by the friction between the porous media and the fluid, we have $\bF_f = - \bF_s$. Since in that case $\dot E \leq 0$, we must necessarily have 
\begin{equation} 
\label{Darcys_friction} 
\dot E = \int_\mathcal{B}   \bF_s \cdot  \left( \bu_s - 
     \bu_f  \right)  \mbox{d}^3 \bx  \leq 0 \, . 
\end{equation} 
If one assumes \eqref{Darcy_law} for the friction, \emph{i.e.},  $\bF_s= \mathbb{K}(\bu_s-\bu_f) $, then 
 $\mathbb{K}$ must be a positive operator, \emph{i.e.}, $\mathbb{K} \bv \cdot \bv \geq 0$, for all $\bv \in \mathbb{R}^3$ and for any point $\bx \in \mathcal{B} $.

\section{Connection with the previously derived models of porous media}

\subsection{The case of a compressible porous media filled  with compressible fluid}
Let us start with connecting to the case considered frequently in the literature, namely, the case of a compressible fluid moving inside a matrix made out of elastic compressible material. In this case, the fluid pressure is no longer a Lagrange multiplier, but has to be found from the identities regarding the internal energy of the fluid as a function of its density. We refer the reader to \cite{landau2013course} for background in classical thermodynamics. If the volume fraction occupied by the fluid is $\phi$, the volume fraction of the elastic matrix is then $1-\phi$.  In the general thermodynamic description, the specific internal energy of the material $e$ is a function of its density $\rho$ and specific entropy $S$, with the pressure being given as $p= \rho  ^2 \pp{e}{\rho}$. This formula is correct whether the thermodynamics effects are considered, \emph{i.e.} $S$ is varying, or ignored, \emph{i.e.} $S=$const. If the effective density of the fluid is $\rho_f$, and its volume fraction  is $\phi$, then the microscopic density of the fluid  is $\bar{\rho}_f=\rho_f/\phi$, so the internal energy of the fluid is a function of $\bar \rho_f$, \emph{i.e.}, $e_f=e_f(\bar \rho_f)$. Similarly, the microscopic density of the solid is $\bar{\rho}_s=\rho_s /(1-\phi)$.  It is natural to assume that the internal energy of the elastic solid depends on both $\bar \rho_s$ and the Finger deformation tensor $b$, $e_s=e_s(\bar \rho_s, b)$. Thus, the physically relevant Lagrangian takes the form 
\begin{equation} 
\label{lagr_compressible} 
\begin{aligned} 
\ell(\bu_f,\bu_s,\rho_f,\rho_s,b,\phi)  =  
\int_{{\mathcal B} }& \left[\frac{1}{2}\rho_f|\bu_f|^2+ \frac{1}{2}\rho_s|\bu_s|^2  - \rho_f  e_f\left(\frac{\rho_f}{\phi}\right) - \rho_s e_s\left(\frac{\rho_s}{1-\phi}, b\right)\right]{\rm d}x.
\end{aligned} 
\end{equation} 
Proceeding as in the derivation of \eqref{eq_gen}, we obtain the following system, written in terms of a general Lagrangian: 
\begin{equation} 
\label{gen_lagr_compressible} 
\left\{
\begin{array}{l}
\vspace{0.2cm}\displaystyle\partial_t \frac{\delta\ell}{\delta\bu_f}+\pounds_{\bu_f}\frac{\delta\ell}{\delta\bu_f} =\rho_f\nabla \frac{\delta\ell}{\delta\rho_f}\\
\vspace{0.2cm}\displaystyle\partial_t \frac{\delta\ell}{\delta\bu_s}+\pounds_{\bu_s}\frac{\delta\ell}{\delta\bu_s} =\rho_s\nabla \frac{\delta\ell}{\delta\rho_s} - \frac{\delta\ell}{\delta b}:\nabla b- 2\operatorname{div} \frac{\delta\ell}{\delta b}\cdot b\\
\vspace{0.2cm}\displaystyle\partial_t \rho_f+ \operatorname{div}(\rho_f \bu_f)=0,\qquad \partial_t \rho_s+ \operatorname{div}(\rho_s \bu_s)=0,\qquad\partial_t b+ \pounds_{\bu_s}b=0\\
\displaystyle\frac{\delta\ell}{\delta\phi}=0\\
\end{array}
\right.
\end{equation} 
When the particular form of the Lagrangian \eqref{lagr_compressible} is assumed, the equations take the form: 
\begin{equation} 
\label{eq_gen_compressible} 
\left\{
\begin{array}{l}
\vspace{0.2cm}\displaystyle \rho_f (\partial_t \bu_f+ \bu_f\cdot \nabla \bu_f) =- \rho_f\nabla \left( e_f + \bar\rho_f \frac{\partial e_f}{\partial\bar\rho_f}\right)= - \phi \nabla \left(\bar\rho_f^2 \frac{\partial e_f}{\partial\bar\rho_f}\right) \\
\vspace{0.2cm}\displaystyle\rho_s( \partial_t \bu_s+ \bu_s\cdot \nabla \bu_s) =- \rho_s\nabla \left( e_s + \bar\rho_s \frac{\partial e_s}{\partial\bar\rho_s}\right) - \rho_s \frac{\partial e_s}{\partial  b}:\nabla b+  2\operatorname{div} \frac{\partial e_s}{\partial  b}\cdot b \\
\vspace{0.2cm}\displaystyle \hspace{3.42cm}= - (1-\phi) \nabla \left(\bar\rho_s^2 \frac{\partial e_s}{\partial\bar\rho_s}\right) +  2\operatorname{div} \frac{\partial e_s}{\partial  b}\cdot b\\
\vspace{0.2cm}\displaystyle\partial_t \rho_f+ \operatorname{div}(\rho_f \bu_f)=0,\qquad \partial_t \rho_s+ \operatorname{div}(\rho_s \bu_s)=0,\qquad\partial_t b+ \pounds_{\bu_s}b=0\\
\displaystyle  \bar\rho_f^2\frac{\partial e_f}{\partial\bar\rho_f} =  \bar\rho_s^2\frac{\partial e_s}{\partial\bar\rho_s}=:p\, \quad 
\mbox{where} \quad 
\displaystyle \bar{\rho}_f:=\frac{\rho_f }{1-\phi}, \quad \bar{\rho}_s=\frac{\rho_s}{\phi} \, . 
\end{array}
\right.
\end{equation} 
The last equation, coming from the variation in $\delta \phi$, states the equality of pressure in both elastic and fluid part of the system. We can transform the system to the following form: 
\begin{equation}\label{classic_porousmedia}
\left\{
\begin{array}{l}
\vspace{0.2cm}\displaystyle \rho_f (\partial_t \bu_f+ \bu_f\cdot \nabla \bu_f) = - \phi \nabla p \\
\vspace{0.2cm}\displaystyle\rho_s( \partial_t \bu_s+ \bu_s\cdot \nabla \bu_s) = - (1-\phi) \nabla p +  \operatorname{div} \sigma_{\rm el}\\
\displaystyle\partial_t \rho_f+ \operatorname{div}(\rho_f \bu_f)=0,\qquad \partial_t \rho_s+ \operatorname{div}(\rho_s \bu_s)=0,\qquad\partial_t b+ \pounds_{\bu_s}b=0\\
\end{array}
\right.
\end{equation}
where
\[
p:= \bar\rho_f^2\frac{\partial e_f}{\partial\bar\rho_f} =  \bar\rho_s^2\frac{\partial e_s}{\partial\bar\rho_s},\qquad \sigma_{\rm el}:= 2\frac{\partial e_s}{\partial  b}\cdot b
\]
Equations similar to \eqref{classic_porousmedia} appear, for example in \cite{ChMo2010}, with additional thermodynamical effects. These thermodynamics effects can be incorporated in our model as well if we allow the energies of the fluid and solid part in the Lagrangian \eqref{lagr_compressible} to depend on the entropies of fluid $S_f$ and solid $S_s$, such as $e_f=e_f(\bar\rho_f, S_f)$ and 
$e_s=e_f(\bar\rho_s, S_s)$, with additional equations for advection of the entropy and heat exchange between the two phases. We shall postpone this discussion of thermal effect for our follow-up work in order not to distract from the main message of the paper.  However, within the framework of this paper, it is worth noting that the internal energies of the fluid and solid are completely separated: the internal energy of the fluid depends only on the internal variables of the fluid, and, correspondingly,  the internal energy of the elastic matrix depends only on the internal variables of the elastic material. The interaction between the terms comes from equality of pressure and follows from the equations of motion; it does not have to be assumed \emph{a priori}. Thus, we believe, our approach is consistent with the classical Lagrangian approach of dealing with the systems with several interacting parts.

\subsection{Compressible media with incompressible fluid}

Let us now connect this description of compressible fluid and solid to the case of incompressible fluid and compressible solid. We shall keep the same variables as in the derivation of \eqref{eq_gen_compressible} to keep the notation consistent, and then show how to connect the resulting equations with \eqref{eq_gen}. 
The difference between the cases of compressible and incompressible fluids comes to two fundamental restrictions: 
\begin{enumerate} 
\item Since the microscopic density of fluid $\bar \rho_f$, also denoted $\rho_f^0$ earlier, is constant, the internal energy of the fluid do not depend on $\bar \rho_f$. 
\item There is an incompressibility condition $\phi = (\phi^0 \circ \bvarphi^{-1})J_{\bvarphi^{-1}}$, equivalent to \eqref{cons_law_fluid}. We remind the reader that $\bvarphi^{-1}(\bx,t)$ is the inverse of the Lagrangian mapping for fluid particles, also known as the back-to-labels map. Physically, this law states that all the fluid in a given microscopic volume of porous media has appeared from its initial source at $t=0$. 
\end{enumerate} 
Note that the incompressibility condition presented above is similar to the conservation of mass in \cite{bedford1979variational} (Eq. (10) taken for the case of fluid only). In spite of this similarity, there is an important difference to keep in mind: in \cite{bedford1979variational}, the conservation law is written for both \emph{compressible} fluid and solid parts. In our case, no additional conservation laws are necessary in the case of compressible fluid and solid, so there is only one incompressibility condition for fluid for the incompressible fluid case, and none for the compressible fluid case. The conservation law for the compressible part in our theory is satisfied automatically, and no extra Lagrange multipliers are necessary.
The action functional \eqref{action_p}, incorporating the constraint with the Lagrange multiplier $p$, rewritten in the new variables, becomes 
\begin{equation} 
\label{action_p2} 
S_p=\int_0^T \left[ \ell(\bu_f,\bu_s, \rho_f, \rho_s, b,\phi) +  \int_\mathcal{B} p \left( \phi - (\phi^0 \circ \bvarphi ^{-1})J_{\bvarphi^{-1}}\right){\rm d}x \right] \mbox{d} t \, .
\end{equation} 
While the method works for an arbitrary Lagrangian, the physically relevant form of the Lagrangian to consider is given by
\begin{equation} 
\label{lagr_incompressible2} 
\ell(\bu_f,\bu_s, \rho_f, \rho_s, b,\phi)= \int_{{\mathcal B}}  \left[\frac{1}{2}\rho_f|\bu_f|^2+ \frac{1}{2}\rho_s|\bu_s|^2 - \rho_s e_s\left(\frac{\rho_s}{1-\phi}, b\right)\right]{\rm d}x.
\end{equation} 
Note that compared to the previous form for compressible fluid case \eqref{lagr_compressible}, the term $ \rho_f  e_f\left( \bar \rho_f \right)$ is now absent from  \eqref{lagr_incompressible2}. Using the identity 
\begin{equation} 
\label{identity_var_phi} 
\delta \left[(\phi^0 \circ \bvarphi ^{-1})J_{\bvarphi^{-1}}\right]= - \operatorname{div}\left((\phi^0 \circ \bvarphi ^{-1})J_{\bvarphi^{-1}} \boldeta_f\right) \, , 
\end{equation} 
we get the following set of equations written for a general Lagrangian $\ell$: 
\begin{equation} 
\label{eq_general_lagr_incompressible} 
\left\{
\begin{array}{l}
\vspace{0.2cm}\displaystyle\partial_t \frac{\delta\ell}{\delta\bu_f}+\pounds_{\bu_f}\frac{\delta\ell}{\delta\bu_f} = \rho_f\nabla \frac{\delta\ell}{\delta\rho_f}  - \phi \nabla p\\
\vspace{0.2cm}\displaystyle\partial_t \frac{\delta\ell}{\delta\bu_s}+\pounds_{\bu_s}\frac{\delta\ell}{\delta\bu_s} =\rho_s\nabla \frac{\delta\ell}{\delta\rho_s} - \frac{\delta\ell}{\delta b}:\nabla b- 2\operatorname{div} \frac{\delta\ell}{\delta b}\cdot b\\
\vspace{0.2cm}\displaystyle\partial_t \rho_f+ \operatorname{div}(\rho_f \bu_f)=0,\qquad \partial_t \rho_s+ \operatorname{div}(\rho_s \bu_s)=0,\qquad\partial_t b+ \pounds_{\bu_s}b=0\\
\displaystyle  \phi = (\phi^0 \circ \varphi_f ^{-1})J_{\varphi_f^{-1}},\qquad \frac{\delta\ell}{\delta\phi}+ p=0.
\end{array}
\right.
\end{equation} 
In the case of the physically relevant Lagrangian \eqref{lagr_incompressible2}, we obtain 
\begin{equation} 
\label{eq_gen_incompressible2} 
\!\!\!\!\left\{
\begin{array}{l}
\vspace{0.2cm}\displaystyle \rho_f (\partial_t \bu_f+ \bu_f\cdot \nabla \bu_f) = - \phi \nabla p \\
\vspace{0.2cm}\displaystyle\rho_s( \partial_t \bu_s+ \bu_s\cdot \nabla \bu_s) =- \rho_s\nabla \left( e_s + \bar\rho_s \frac{\partial e_s}{\partial\bar\rho_s}\right) - \rho_s \frac{\partial e_s}{\partial  b}:\nabla b+  2\operatorname{div} \frac{\partial e_s}{\partial  b}\cdot b \\
\vspace{0.2cm}\displaystyle \hspace{3.42cm}  = - (1-\phi) \nabla \left(\bar\rho_s^2 \frac{\partial e_s}{\partial\bar\rho_s}\right) +  2\operatorname{div} \frac{\partial e_s}{\partial  b}\cdot b\\
\vspace{0.2cm}\displaystyle\partial_t \rho_f+ \operatorname{div}(\rho_f \bu_f)=0,\qquad \partial_t \rho_s+ \operatorname{div}(\rho_s \bu_s)=0,\qquad\partial_t b+ \pounds_{\bu_s}b=0\\
\displaystyle \partial_t \phi+ \operatorname{div}(\phi \bu_f)=0,\qquad \bar\rho_s^2\frac{\partial e_s}{\partial\bar\rho_s}=p.
\end{array}
\right.
\end{equation} 
Equations $\partial_t \rho_f+ \operatorname{div}(\rho_f \bu_f)=0$ and $ \partial_t \phi+ \operatorname{div}(\phi \bu_f)=0$ imply that $ \rho_f= \bar\rho_f \phi$ with $\bar\rho_f$ a constant. Note that the last equation of \eqref{eq_gen_incompressible2}, states that the thermodynamic pressure in the solid, defined through the derivatives of the internal energy function $e_s$, is equal to the Lagrange multiplier $p$. Thus, physically, the Lagrange multiplier $p$ is equal to the pressure inside the solid, so it also acquires the physical meaning of the pressure in the fluid. However, that physical meaning is elucidated only \emph{after} the equations of motion \eqref{eq_gen_incompressible2} are derived and cannot be inferred \emph{a priori}. 

A quick calculation shows that the system \eqref{eq_gen_incompressible2} is equivalent to the equations \eqref{eq_gen} derived earlier, under the change of variables 
\begin{equation} 
\label{change_var_equiv} 
g=\phi, \quad c(b)=\rho_s,\quad v= \frac{1}{\rho_s}- \frac{1}{\bar\rho_s} \, . 
\end{equation}
That equivalence is proved by assuming the internal energy of the solid in the form 
\begin{equation} 
\label{int_energy_connect} 
V(b,v)= \rho_s(b) \emph{e}_s \left( \bar \rho_s , b \right)\, , \quad \bar\rho_s:= \frac{\rho_s(b)}{1- \rho  _s (b)v} \, .
\end{equation} 
Substitution of that expression for the internal energy of the solid into \eqref{eq_gen} gives \eqref{eq_gen_incompressible2}. We believe that such calculation is useful since it connects our earlier derivation \eqref{eq_gen} with the information on the compressible case, and also elucidates the nature of the variable $\phi$. It is useful to recall the quote from \cite{wilmanski2006few}  mentioned in the Introduction, where the nature of this variable was suggested to preclude the existence of a variational principle. Our theory presented here shows that the variable describing the fluid content has to be considered carefully in the variational principle \eqref{action_p2}, or, equivalently, in \eqref{action_p}  earlier, as a constraint through the geometric variational formulation presented here.  The understanding of the role of this variable, we believe, is the key to the derivation of the variational principle for porous media, and was perhaps the source of difficulty in explaining the incompressible fluid case in previous works. 

The physical meaning of $v$  becomes clear from the last formula of \eqref{change_var_equiv}. Indeed, choose $m_s$ to be a given mass of elastic solid, then $\frac{m_s}{\rho_s}$ is the volume of occupied by the porous elastic solid, and $\frac{m_s}{\bar\rho_s} $ is the volume occupied by the (imaginary) elastic solid without any porosity. Thus, the quantity $m_s  \left( \frac{1}{\rho_s}- \frac{1}{\bar\rho_s} \right)$ is the  volume occupied by the fluid per unit mass of the solid, and therefore the quantity $v= \frac{1}{\rho_s}- \frac{1}{\bar\rho_s} $ is the physical meaning of specific volume of the fluid's content, measured per unit mass of the elastic solid.

\rem{ 
\begin{theorem} This system is equivalent to ours, by using the relations
\[
g=\phi, \quad c(b)=\rho_s,\quad v= \frac{1}{\rho_s}- \frac{1}{\bar\rho_s}
\]
where we choose $c_0=1$ and we assume $\varrho_{s0}$ is a constant.
\end{theorem}
\begin{proof} This is a direct computation: we compute our system with $V(b,v)$ given by
\[
V(b,v)= \rho_s(b) E_s \left( \bar\rho_s= \frac{\rho_s(b)}{1- \rho  _s (b)v} , b \right)
\]
and we get the system above.
\end{proof}
} 

\medskip

\rem{ 
The advantage of this formulation is that it is closer to the usual treatment in the literature and it is more clear about the Lagrange multiplier $p$ which is literally given by the 
usual definition of pressure $p=\bar\rho_s^2\frac{\partial E_s}{\partial\bar\rho_s}$ via the $ \delta \phi $-equation. It is certainly easier to include thermodynamics too.
} 

\section{Linear stability analysis}

\label{sec:lin_stab}

\subsection{Derivation of the linearized equations of motion} 

We linearize equations \eqref{expressions_explicit} about the equilibrium state
\begin{equation} 
\label{equilibrium_state} 
(\bu_f,\bu_s,\rho_s,b,g,v,p)=(\mathbf{0},\mathbf{0},\rho_s^0, b_0, g_0,v_0,p_0)\,,
\end{equation} 
where each component on the right-hand side of \eqref{equilibrium_state} with a subscript $0$ is a constant.
The equilibrium condition reads
\begin{equation}\label{equilibrium_condition}
\left.\frac{\partial V}{\partial v}\right|_0=p_0 c_0\,.
\end{equation}
where $F|_0$ denotes the value of a function $F$ taken at the equilibrium \eqref{equilibrium_state}.
We consider the potential $V(b,v)$ to be general and assume, for simplicity, an unstressed state $b_0 =\id$ and $p_0=0$. Throughout this section, we shall assume friction forces of the form \eqref{Darcy_law} with a given constant general permeability tensor $\mathbb{K}$. For simplicity of computations, we will eventually further assume isotropic and uniform media, so the permeability tensor $\mathbb{K}$ will be taken proportional to a unity matrix.

\paragraph{Notation.} In this chapter on linearization, we denote the value of a variable $f$ evaluated at the equilibrium with the index $0$, \emph{i.e.}, $f_0$. The spatiotemporal deviation from the equilibrium is then denoted as $\de f(\bx,t) \simeq f(\bx,t)-f_0(\bx,t)$, with $\delta f$ assumed small.  Note that this is the same notation $\delta$ as for the variations used in the previous chapter. We hope that no confusion arises due to that clash of notation.  

\paragraph{Expression of the stress tensor.} The full stress tensor computed from \eqref{sigma_p} is
\[
\sigma_p= \sigma_{\rm el} + \frac{c_0 vp}{J}\id=\sigma_{\rm el} + cvp \id=\sigma_{\rm el} + gp \id,\qquad J= \sqrt{\det b}\,,
\]
where for $c(b)= c_0/J$, we used
\[
\frac{\partial c}{\partial b}= - \frac{c_0}{2J} b^{-1}
\]
and where
\[
\sigma_{\rm el}= 2 \frac{\partial V}{\partial b}\cdot b
\]
is the elastic stress tensor associated to the potential $V$.
The linearization of the full stress tensor is
\begin{equation}\label{delta_sigma_p}
\delta\sigma_p= \delta \sigma_{\rm el} + g_0 \delta p\id,
\end{equation}
where we recall that we chose $p_0=0$ and that $b_0=\id$, so $J|_0 = 1$. The linearization of the elastic stress tensor is written as
\begin{equation}\label{delta_sigma_el}
\delta \sigma_{\rm el}= \left.\frac{\partial\sigma_{\rm el}}{\partial b }\right|_0:\delta b + \left.\frac{\partial\sigma_{\rm el}}{\partial v }\right|_0\delta v= 2 \left.\frac{\partial ^2V}{\partial b^2}\right|_0:\delta b+ 2\left. \frac{\partial V}{\partial b}\right|_0\cdot \delta b
+ 2 \left.\pp{^2V}{b\partial v}\right|_0 \delta v.
\end{equation}

\paragraph{Linearization.} 
The system \eqref{expressions_explicit} is linearized as follows: 
\rem{ 
\begin{equation}
\label{expressions_explicit_V} 
\left\{
\begin{array}{l}
\displaystyle\vspace{0.2cm}\rho_f g (\partial_t \bu_f+ \bu_f\cdot\nabla \bu_f )  = - g\nabla  p + \mathbb{K}(\bu_s - \bu _f)\\
\displaystyle\vspace{0.2cm}\rho_s (\partial_t \bu_s +\bu_s\cdot\nabla \bu_s) =g\nabla p \!+\! \nabla \left( V\! - \! \pp{V}{v}v \right) + \operatorname{div} \sigma_p + \mathbb{K}(\bu_f - \bu _s)\\
\displaystyle\vspace{0.2cm} \pp{V}{v}=  pc(b),\qquad g= c(b)v\\
\vspace{0.2cm}\partial_tg + \operatorname{div}(g\bu_f)=0,\qquad \partial_t\rho_s+\operatorname{div}(\rho_s\bu_s)=0,\qquad \partial_tb+ \pounds_{\bu_s}b=0.
\end{array}\right.
\end{equation}
and its linearization is
} 
\begin{equation}
\label{linearized_equations} 
\hspace{-0.3cm}\left\{
\begin{array}{l}
\displaystyle\vspace{0.2cm}g_0\rho_f\partial_t \delta \bu_f = - g_0 \nabla  \delta p + \mathbb{K} (\delta \bu_s- \delta \bu_f)\\
\displaystyle\vspace{0.2cm}\rho_s^0 \partial_t\delta  \bu_s = 
\nabla \left(\left.\pp{V}{b}\right|_0:\delta b \right) + \operatorname{div} \delta \sigma_p  + \mathbb{K} (\delta \bu_f- \delta \bu_s)\\
\displaystyle\vspace{0.2cm} \left.\pp{^2V}{v^2}\right|_0 \delta v + \left.\pp{^2V}{v \partial b}\right|_0:\delta b =  c_0\delta p \,,\qquad \delta g= -\frac{c_0}{2} \tr{\delta b}v_0 +  c_0\delta v\\
\vspace{0.2cm}\partial_t\delta g + \operatorname{div}(g_0\delta\bu_f)=0\,,\qquad \partial_t\delta\rho_s+\operatorname{div}(\rho_s^0\delta\bu_s)=0\,,\\ 
\vspace{0.2cm} \partial_t\delta b- 2\operatorname{Def}\delta \bu_s =0\, , \quad 
\operatorname{Def} \delta \bu_s := \frac{1}{2} \left( \nabla \delta \bu_s + \left[ \nabla \delta \bu_s \right]^T \right) \, . 
\end{array}\right.
\end{equation}
To get the linearized balance of elastic momentum, we used the fact that the linearization of the term $\nabla (V- v \pp{V}{v} ) = \nabla (V- pg)$ for $p_0=0$ in the second equation of \eqref{expressions_explicit} is computed as 
\begin{equation}\label{intermediate_step}
\delta \nabla \left( V-p g \right) = 
\nabla \Big(  \underbrace{\pp{V}{v}\Big|_0}_{=p_0 c_0=0}  \delta v +   \pp{V}{b}\Big|_0 :\delta b \Big) - g_0 \nabla \delta p\, .
\end{equation} 
The last term in \eqref{intermediate_step} then cancels with the linearization of the first term on the right hand side of \eqref{expressions_explicit}, thus yielding the second equation in \eqref{linearized_equations}.

To get the last equation in \eqref{linearized_equations} we used that the linearization of the Lie derivative $\pounds_{\bu_s}b$ at $\bu_{s,0}=0$ and $b_0=\id$ is $- 2\operatorname{Def}\delta \bu_s$ as a direct computation using \eqref{Lie_der_b} shows.

For the linearized equations, we shall only need the coefficients of the linear and the quadratic expansions of the potential $V(b,v)$ about the equilibrium. We thus define the coefficients: 
\begin{equation} 
\sigma_0 = \left. \pp{V}{b} \right|_0 \,, 
\qquad 
\zeta =\frac{v_0}{c_0} \left.   \pp{^2 V}{v^2} \right|_0 \, , 
\qquad 
\mathbb{C} = \left.    \frac{\partial^2 V}{ \partial b^2 } \right|_0\, ,
\qquad 
\mathbb{D} = \left.  \frac{\partial^2 V}{\partial v \partial b } \right|_0 \, . 
\label{variables_def} 
\end{equation}
The coefficient $\zeta$, from its definition, has the order of magnitude of the bulk modulus of the microscopic material itself, although it can depend on the pores concentration and their arrangement in the matrix.
Using this, the potential energy of the elastic deformation $V(b,v)$ about the equilibrium, up to the second order in deviations from equilibrium, and assuming $V(b_0,v_0)=0$, is represented as 
\begin{equation}\label{expansion_potential} 
\begin{aligned} 
V(b,v) &\simeq \sigma_0:(b-b_0) + \frac{1}{2}(b-b_0) :\mathbb{C}: (b-b_0) + \frac{c_0\zeta}{2v_0}(v-v_0)^2 +\mathbb{D}:(b-b_0)(v-v_0) \, .
\end{aligned} 
\end{equation}
From \eqref{variables_def}  and \eqref{delta_sigma_el}, we have
\begin{equation}\label{linearized_sigma_p} 
\delta \sigma _p= \delta \sigma _{\rm el}+  g_0 \delta p\id= 2\mathbb{C}: \delta b + 2 \sigma _0 \cdot \delta b + 2 \mathbb{D} \delta v +  g_0 \delta p\id.
\end{equation} 
The first term identifies the Hooke law connecting the linearized stress and linearized strain $\epsilon$ as follows
\begin{equation} 
\label{sigma_1_eq}
\sigma_1 := 2 \left.    \frac{\partial^2 V}{ \partial b^2 } \right|_0:\delta b =  2\mathbb{C}: \delta b = 4 \mathbb{C}: \epsilon \, , \qquad \epsilon := \frac12 \delta b \simeq \frac{1}{2} (b-b_0) \, ,
\end{equation} 
where the definition of $\epsilon$ above is understood as a linearization of $b$ about the equilibrium. We have intentionally denoted this linearized part of Finger tensor as $\epsilon$ since it happens to be exactly the standard linear strain used in elasticity, see \eqref{Hookeslaw} below.

We shall now assume an isotropic and uniform material, which will be the case of study for the remainder of the paper.   Then, the tensor $\mathbb{C}$ in \eqref{sigma_1_eq} has only two independent coefficients, and \eqref{sigma_1_eq} becomes the familiar Hooke law for isotropic uniform materials, \emph{i.e.},
\begin{equation}
\label{Hookeslaw}
\sigma_1(\epsilon) = 4 \mathbb{C}: \epsilon= 2G\epsilon + \Lambda \tr{\epsilon}\, \id,
\end{equation}
where $\Lambda$ and $G$ are well known as Lam\'{e} parameters for isotropic materials in continuum mechanics \footnote{Sometimes Lam\'e coefficients are denoted $\lambda$ and $\mu$. We will avoid that notation since it clashes with the notation used here, where $\lambda$ denotes the growth rate of the disturbances as defined in \eqref{disturbances_def}, and $\mu$ characterizing the residual stress according to \eqref{mu_xi_def}.  We hope no confusion arises from our use of notation for Lam\'{e} coefficients. }. 
Furthermore, the tensors $\sigma_0$ and $\mathbb{D}$ in \eqref{variables_def} are proportional to the unit tensor, and it is convenient to express them as follows:
\begin{equation} 
\sigma_0 = \frac12 g_0 \mu \, \id \, , \quad 
\mathbb{D}=\frac12 {c_0\xi} \, \id\, , \mbox{ with } \mu, \xi={\rm const. }
\label{mu_xi_def}
\end{equation} 
The constants $\mu$ and $\xi$, defined above, as well as coefficient $\zeta$ defined by \eqref{variables_def}, 
have the dimension of Young's modulus, \emph{i.e.}, pressure. With all these assumptions \eqref{linearized_sigma_p} becomes 
\begin{equation}\label{linearized_sigmaell}
\delta \sigma_p= \frac{\Lambda}{2}\tr{\delta b} \id + (G + g_0\mu) \delta b + c_0\xi\delta v\,\id  + g_0 \delta p\id \,.
\end{equation}
\rem{$\tilde\lambda= \tilde\kappa- \frac{2}{3}\tilde \mu= \frac{E\nu}{(1+\nu)(1-2\nu)}$, which exactly recovers \eqref{sigma_dist_lambda}}

\paragraph{Linear stability.} We now set
\begin{equation} 
\label{disturbances_def} 
\begin{aligned}
\delta \bu_s &= \lambda \bv e^{\lambda t + i \bk \cdot \bx}\,,  
& \delta \bu_f &= \lambda \bu e^{\lambda t + i \bk \cdot \bx}\,, 
& \delta \rho_s&= \rho_{s,1} e^{\lambda t + i \bk \cdot \bx}\,, 
& \delta b &=b_ 1 e^{\lambda t + i \bk \cdot \bx}\,, 
\\ 
 \delta g&= g_1 e^{\lambda t + i \bk \cdot \bx}\,,
 & \delta v &=v_ 1 e^{\lambda t + i \bk \cdot \bx}\,, &  \delta p &=p_ 1 e^{\lambda t + i \bk \cdot \bx}\,. & & 
\end{aligned}
\end{equation} 
and equations \eqref{linearized_equations} become
\rem{
\[
(\operatorname{div}\delta \sigma_p)_k= \left(p_1 v_0 c_0 i \bk_k + 2 \left.\frac{\partial ^2V}{\partial b^{ki}\partial b^{mn}}\right|_0: b_1^{mn} i\bk_i-  \left.\frac{\partial ^2V}{\partial b^{ki}\partial v}\right|_0 v_1 i \bk_i+ 2 \left.\frac{\partial V}{\partial b^{kj}}\right|_0  b^{ij}_1 i \bk _i\right)e^{\lambda t + i \bk \cdot\bx}\,.
\]}
\begin{equation}
\label{expressions_linearized} 
\left\{
\begin{array}{l}
\displaystyle\vspace{0.2cm}g_0\rho_f\lambda^2 \bu = - g_0 p_1 i \bk + \lambda\mathbb{K} (\bv-  \bu)\\
\displaystyle\vspace{0.2cm}\rho_s^0 \lambda^2 \bv =  \frac12({\Lambda}+{g_0\mu})\tr{b_1} i \bk + i (G + g_0\mu) b_1\cdot \bk + i \bk(g_0 p_1 + c_0\xi v_1) + \lambda\mathbb{K} (\bu -  \bv)\\
\displaystyle\vspace{0.2cm} \frac{\zeta}{v_0} {v_1} + \frac{\xi}{2} \tr{b_1}=   p_1 \,,\qquad g_1= -\frac{c_0}{2} \tr{b_1}v_0 +  c_0 v_1\,,\\
\displaystyle\vspace{0.2cm}g_1+ g_0 i ( \bu \cdot \bk)=0\,,\qquad \rho_{s,1}+ \rho_s^0 i ( \bv \cdot \bk)=0\,,\\
\displaystyle b_1 -   i ( \bv \otimes  \bk +  \bk \otimes  \bv)=0\,.
\end{array}\right.
\end{equation}

By using the expression $c(b)= c_0/J$ and the last equation in \eqref{expressions_linearized}, we get $c_1= - c_0\tr{b_1}/2= - c_0 i (\bv\cdot \bk)$ so, from the constraint $g=c v$ we have
\begin{equation}\label{g_1}
g_1 = c_0 v_1 + c_1 v_0 = c_0 ( v_1- i v_0 (\bv \cdot \bk) ) \, .
\end{equation}
From the linearized continuity equation for $g$ in \eqref{expressions_linearized} we get $g_1 = - i g_0  ( \bu \cdot \bk)$, so combining this with \eqref{g_1} we obtain the expression of $v_1$ as
\begin{equation} 
\label{v_1} 
v_1 =  i v_0 \left( \bv \cdot \bk - \bu \cdot \bk \right),
\end{equation} 
\rem{\[\sigma_{\rm el}= 2(\sigma_1 + \sigma_0 \cdot \delta b)\]}
and then, substituting this result into the third equation of \eqref{expressions_linearized}, we deduce \begin{equation} 
\label{pressure_expr} 
p_1  =i \left( \left( \zeta+ \xi \right)  (\bv \cdot \bk)  - \zeta(\bu \cdot \bk) \right)  \, . 
\end{equation}
From \eqref{v_1} and \eqref{pressure_expr}, we get
\begin{equation}
g_0 p_1+c_0 \xi v_1 = ig_0\left((\zeta + 2\xi)(\bv \cdot \bk) - (\zeta+\xi)(\bu \cdot \bk)\right),
\end{equation}
and the first two equations in \eqref{expressions_linearized} become
\begin{equation}
\label{expressions_linearized2} 
\hspace{-0.3cm}
\left\{
\begin{array}{l}
\displaystyle\vspace{0.2cm}g_0\rho_f\lambda^2 \bu = g_0\bk \left( \left( \zeta+ \xi \right)  (\bv \cdot \bk)  - \zeta(\bu \cdot \bk) \right) + \lambda\mathbb{K} (\bv-  \bu)\\
\displaystyle\vspace{0.2cm}\rho_s^0 \lambda^2 \bv = - (\Lambda+g_0\mu) (\bv\cdot\bk)  \bk - (G+g_0\mu) ( \bv |\bk|^2+ \bk ( \bv\cdot \bk)) \\
\qquad \qquad + g_0\bk[(\zeta + \xi)(\bu \cdot \bk) - (\zeta + 2\xi)(\bv \cdot \bk)]
+\lambda\mathbb{K} (\bu -  \bv) \, . 
\end{array}\right.
\end{equation}
\rem{ 
we can define new coefficients
\begin{equation}
\label{effective_pseudolame_coef}
\hat{\Lambda}:=\Lambda+g_0(\mu+\xi),\quad \hat{G}:=G+g_0\mu,
\end{equation} so that our system \eqref{expressions_linearized2} will be simplified to
\begin{equation}
\label{expressions_linearized3} 
\hspace{-0.3cm}\left\{
\begin{array}{l}
\displaystyle\vspace{0.2cm}g_0\rho_f\lambda^2 \bu = g_0\bk \left[ \left( \zeta+ \xi \right)  (\bv \cdot \bk)  - \zeta(\bu \cdot \bk) \right] + \lambda\mathbb{K} (\bv-  \bu)\\
\begin{aligned}
\displaystyle\vspace{0.2cm}\rho_s^0 \lambda^2 \bv = & - \hat{\Lambda} (\bv\cdot\bk)  \bk - \hat{G} ( \bv |\bk|^2+ \bk ( \bv\cdot \bk)) \\& + g_0\bk[(\zeta + \xi)(\bu \cdot \bk) - (\zeta + \xi)(\bv \cdot \bk)]
+\lambda\mathbb{K} (\bu -  \bv),
\end{aligned}
\end{array}\right.
\end{equation}
The equation satisfied by the eigenvalue $\lambda$ is given by vanishing determinant of the $6 \times 6$ matrix $\mathbb{S}$ for the components $(\bu,\bv)^T$: 
\begin{equation} 
\label{matr_def} 
\begin{aligned} 
\operatorname{det}\mathbb{S}=0, \quad \mathbb{S}&=
\left[ 
\left( 
\begin{array}{cc} 
\lambda^2 \rho_f g_0 \id & 0 
\\ 
0 & \lambda^2 \rho_s^0 \id 
\end{array} 
\right) 
+ 
\lambda 
\left( 
\begin{array}{cc} 
\mathbb{K} & - \mathbb{K} 
\\ 
-\mathbb{K} &  \mathbb{K} 
\end{array} 
\right) \right. 
\\
& \qquad \qquad \left. 
+ 
\left( 
\begin{array}{cc} 
g_0 \zeta \mathbb{A}  & -g_0 ( \zeta+\xi) \mathbb{A} 
\\ 
- g_0 {(\zeta+\xi)} \mathbb{A}   & 
g_0 ( \zeta+\xi ) \mathbb{A}  + \mathbb{B} 
\end{array} 
\right) 
\right], 
\end{aligned} 
\end{equation} 
} 

Let us now compute the dispersion relation explicitly for the case when the dissipation in the media is isotropic, so $\mathbb{K}=\beta \id$ for some $\beta>0$. In that case, we obtain the dispersion relation $\operatorname{det}\mathbb{S}=0$ with the matrix $\mathbb{S}$ of the form 
\begin{equation} 
\label{matr_def_isotropic} 
\begin{aligned}  
\mathbb{S} & =
\left[
\begin{array}{cc} 
\lambda^2 \rho_f g_0 \id & 0 
\\ 
0 & \lambda^2 \rho_s^0 
\end{array} 
\right] 
+ 
\lambda 
\beta \left[ 
\begin{array}{cc} 
\id   & - \id 
\\ 
-\id  & \id 
\end{array} 
\right] 
+ 
\left[ 
\begin{array}{cc} 
g_0 \zeta \mathbb{A}  & -g_0 ( \zeta+\xi) \mathbb{A} 
\\ 
- g_0 {(\zeta+\xi)} \mathbb{A}   &
g_0 ( \zeta+\xi ) \mathbb{A} + \mathbb{B} 
\end{array} 
\right]\,,
\end{aligned} 
\end{equation} 
where 
\begin{equation} 
\label{def_A_B}
\mathbb{A}:=\bk \otimes \bk \, , \quad 
\mathbb{B}:= ( \Lambda + G+ g_0 (2 \mu+ \xi)) \mathbb{A} + (G+g_0 \mu) |\bk|^2\, \id\,.
\end{equation}

\begin{remark}[On formal equivalence of Lam\'e coefficients] 
{\rm  
One can notice that in \eqref{matr_def_isotropic}, $G$, $\Lambda$ and $\mu$ only enter in combinations $G+g_0 \mu$ and $\Lambda + g_0 \mu$. Therefore, the acoustic properties of the media with Lam\'e coefficients $G$, $\Lambda$ and with $\left.\frac{\partial V}{\partial b}\right|_0=\frac{1}{2}\mu g_0\neq 0$ are the same as the acoustic properties of the media with the Lam\'e coefficients replaced by the shifted values 
\begin{equation} 
G \rightarrow G+ g_0 \mu \, , \quad \Lambda \rightarrow \Lambda + g_0 \mu, \quad \mbox{and with} 
\quad  \left.\frac{\partial V}{\partial b}\right|_0=0 \, . 
\label{Lame_shift} 
\end{equation} 
Since $G>0$ for consistency of the media, we must have 
\begin{equation} 
\label{G_consitency} 
G+ g_0 \mu>0 \, . 
\end{equation} }
\end{remark} 
Or, more generally, two medias with $G_i,\Lambda_i$ and $\left.\frac{\partial V}{\partial b}\right|_0=\frac{1}{2}\mu_i g_0$, $i=1,2$ are equivalent if $G_1+g_0 \mu_1= G_2+g_0 \mu_2$ and $\Lambda_1+g_0 \mu_1= \Lambda_2+g_0 \mu_2$.

\paragraph{Nondimensionalization.} It is convenient to define the dimensionless growth rates and wavenumbers by choosing the length scales $L$ and time scales $T$: 
\begin{equation} 
\label{dim_var_def} 
\lambda_*= T \lambda\, , \quad \bk_*=L \bk \, . 
\end{equation} 
and $\delta = \rho_f g_0 /\rho_s^0$ is the ratio between the effective equilibrium density of the fluid and the equilibrium density of the elastic material. 

Let us define the following dimensionless matrices 
\begin{equation} 
 \mathbb{A}_*=\bk_* \otimes \bk_* \, , \quad 
 \mathbb{B}_*= \left(1+\frac{\Lambda + g_0(2 \mu+ \xi)}{G}\right) \mathbb{A}_* +  \left(1+\frac{ g_0\mu}{G}\right) |\bk_*|^2\,\id  \, . 
\label{dimensionless_param_def} 
\end{equation} 
Then, dividing $\mathbb{S}$ defined by \eqref{matr_def_isotropic} by $\rho_s^0$, 
we obtain the following dimensionless dispersion matrix defining the equation for nonlinear eigenvalues (growth rates) $\lambda_*$
\begin{equation}  
\begin{aligned} 
\mathbb{S}_*&=
\lambda_*^2 
\left[ 
\begin{array}{cc} 
\delta \id & 0 
\\
0 & \id 
\end{array} 
\right] 
+ \lambda_* \frac{ \beta T}{\rho_s^0} 
\left[ 
\begin{array}{cc} 
\id & -\id
\\
-\id & \id 
\end{array} 
\right] 
\\ 
&\qquad\qquad +
 \frac{g_0  T^2}{\rho_s^0 L^2} 
 \left[ 
\begin{array}{cc} 
\zeta \mathbb{A}_* & - (\zeta+\xi) \mathbb{A}_*
\\
-{(\zeta+\xi)} \mathbb{A}_* & (\zeta+\xi) \mathbb{A}_*
\end{array} 
\right] 
+ \frac{G T^2}{ \rho_s^0 L^2} 
 \left[ 
\begin{array}{cc} 
0  &  0
\\
0 &  \mathbb{B}_*
\end{array} 
\right].
\end{aligned} 
\label{S_non_dim_def} 
\end{equation}
We are free to choose the time and length scales $T$ and $L$, and we choose them in such a way that the coefficients of $\lambda_*$ (friction term) and the last term in \eqref{S_non_dim_def} are equal to unity. This corresponds to choosing 
\begin{equation} 
T=\frac{\rho_s^0}{\beta}, \quad L=T \sqrt{\frac{G}{ \rho_s^0}} \, . 
\label{T_L_choice} 
\end{equation} 
Physically, $T$ is the typical relaxation time in the porous media; $L$ is the distance the elastic sound waves in the 
matrix filled with fluid propagate during that relaxation time. 
We then define the dimensionless quantities 
\begin{equation} 
\label{zeta_star} 
\zeta_*= \frac{ g_0 T^2}{\rho_s^0 L^2}\zeta =g_0 \frac{\zeta}{G} \, , \quad 
\xi_* = \frac{g_0 T^2}{\rho_s^0 L^2} \xi =g_0 \frac{\xi}{G} \, ,
\quad \mu_*=
\frac{g_0\mu}{G}
\, . 
\end{equation} 
With these definitions, the nondimensionalized dispersion matrix takes the form: 
\begin{equation} 
\label{S_non_dim_simplified} 
\mathbb{S}_*=
\left[ 
\begin{array}{cc} 
\id( \delta \lambda_*^2 + \lambda_*) & - \id \lambda_* 
\\
- \id \lambda_*  & \id (\lambda_*^2 + \lambda_*) 
\end{array} 
\right]    +
 \left[ 
\begin{array}{cc} 
\zeta_* \mathbb{A}_* & -(\zeta_* +\xi_*) \mathbb{A}_*
\\
-{(\zeta_*+\xi_*)} \mathbb{A}_* & (\zeta_* +\xi_*) \mathbb{A}_* +\mathbb{B}_*
\end{array} 
\right]\, .
\end{equation} 
Equation $\operatorname{\det} \mathbb{S}_*=0$ defines a 12-th order polynomial in $\lambda_*$, and thus there are exactly 12 roots $\lambda_*=\lambda_*(\bk_*)$ in the complex plane. We now show that given $\bk_*$, all these roots can be computed as $S$- and $P$-waves by considering subspaces parallel and orthogonal to a given $\bk_*$. 

\subsection{$S$-waves}

Let us consider the case in which  $(\bu,\bv) \perp \mathbf{k}_*$. Since $\mathbb{A}_*=\bk_* \otimes \bk_*$, we have $\mathbb{A}_*\bu= \mathbb{A}_*\bv=\mathbf{0}$. In other words, we only consider the displacements orthogonal to the wave vector $\bk_*$, which is exactly the definition of an $S$-wave. We can set $\bu_\perp$ and $\bv_\perp$ to be parallel to a given vector $\bxi$ in the plane $\bk_*^\perp$, \emph{i.e.}, 
$\bu_\perp=u \bxi$ and $\bv_\perp=v \bxi$. The eigenvalues have multiplicity 1 and are computed from the $2 \times 2$ matrix: 
\begin{equation} 
\label{S_non_dim_simplified_S} 
\begin{aligned} 
\mathbb{S}_{*,s}&=
\left[ 
\begin{array}{cc} 
 ( \delta \lambda_*^2 + \lambda_*) & -  \lambda_* 
\\
-  \lambda_*  &  (\lambda_*^2 + \lambda_*) 
\end{array} 
\right] 
  +
|\bk_*|^2  \left[ 
\begin{array}{cc} 
 0 & 0
\\
0 & 1+\mu_*
\end{array} 
\right] \, . 
\end{aligned}
\end{equation} 
Since the space $\bk_*^\perp$ is two-dimensional, all the eigenvalues of $\operatorname{\det} \mathbb{S}_*=0$ given by 
\eqref{S_non_dim_simplified} with $\bu \parallel \bxi$ and $\bv \parallel \bxi$ have multiplicity 2. The equation $\operatorname{\det} \mathbb{S}_{*,s}=0$ given by \eqref{S_non_dim_simplified_S} defines a fourth-order polynomial having 4 roots. Because of the multiplicity 2 of the $S$-waves, the total number of roots for $S$-waves is 8. 

The condition $\operatorname{\det} \mathbb{S}_{*,s}=0$ gives either $\lambda_*=0$, or $\lambda_*$ satisfying the following cubic equation: 
\begin{equation} 
\label{s_wave_disp} 
\delta \lambda_*^3 + \lambda_*^2 (1+ \delta) + \lambda_* k_*^2  (1+ \mu_*) \delta +  k_*^2 (1+\mu_*)=0\, , \quad k_*:=\|\bk_*\| \, . 
\end{equation}  
By Routh-Hurwitz' criterion, the polynomial $s^3 + a_2 s^2 + a_1 s + a_0$ is stable if $a_2 a_1>a_0$. Thus, \eqref{s_wave_disp} is stable, \emph{i.e.}, for any real $k_*$, ${\rm Re} \, \lambda_*<0$, as long as $\delta>0$ (which is natural since $\delta$ is the ratio of densities), and $ \mu_*>-1$. Note that this is exactly the requirement \eqref{G_consitency} for consistency of the media. 

Alternatively, instead of the dispersion relation $\lambda_*=\lambda_*(\bk_*)$, it is common in the literature to compute the attenuation of harmonic signals in porous media, in other words, $k_*(\omega_*)$ when $\lambda_*=i \omega_*$, with $\omega_* \in \mathbb{R}$ being the frequency of forcing. In that case, from \eqref{s_wave_disp} we obtain 
\begin{equation} 
\label{s_wave_attenuation} 
k_*(\omega_*)=\pm \omega_* \sqrt{ \frac{1+ \delta + i \delta \omega_*}{(1+ \mu_*)(1+ i \delta \omega_*)}  }  \, . 
\end{equation}  
As one can see, for $\delta>0$ and $\mu_*>-1$, ${\rm Im}\, k_* \rightarrow 0 $ when $\omega_* \rightarrow 0$, so the attenuation of low-frequency waves decreases with decreasing frequency, which is physically reasonable.  If one considers propagation of waves for $x>0$, one needs to choose the sign in the equation for $k_*(\omega_*)$ in such a way that ${\rm Im}\,k_*(\omega_*)>0$, so the waves will be decaying as $x \rightarrow \infty$.

\subsection{$P$-waves}

Consider the case $(\bu,\bv) \parallel \mathbf{k}$. In other words, we consider the disturbances parallel to the wave vector $\bk$, which is the definition of a $P$-wave. Then $\mathbb{A}_* \bu = (\bk _*\cdot \bu) \bk_*=|\bk_*|^2\bu$, and $\mathbb{A} _*\bv = (\bk_* \cdot \bv) \bk_*=|\bk_*|^2\bv$, and the dispersion relation $\operatorname{det} \mathbb{S}_*=0$ takes the form $\operatorname{det} \mathbb{S}_{*,p}=0$ for the $2\times2$ matrix
\begin{equation} 
\label{disp_P} 
\begin{aligned} 
\mathbb{S}_{*,p}&=
\left[ 
\begin{array}{cc} 
\delta \lambda_*^2 + \lambda_* & -  \lambda_* 
\\
-  \lambda_*  &  \lambda_*^2 + \lambda_*
\end{array} 
\right]
  +
k_*^2  \left[ 
\begin{array}{cc} 
\zeta_*  &-( \zeta_* +\xi_*)
\\
-{(\zeta_*+\xi_*)}   & \zeta_* +2\xi_*  +Z 
\end{array} 
\right] \, ,
\end{aligned} 
\end{equation} 
where we defined for shortness 
\begin{equation} 
\label{combine_params} 
k_*:=   \| \bk \| \, , \quad 
Z:= 2 + \frac{\Lambda}{G} + 3 \mu_* 
\end{equation} 
and we used $\frac{g_0\xi}{G}=\xi_*$ by \eqref{T_L_choice} and \eqref{zeta_star}.
We rewrite this dispersion relation as
\begin{equation} 
\label{lambda_P_non_dim} 
{\rm det} \left[ \begin{array}{cc}
   \delta \lambda_*^2 + \lambda_* + k_*^2 \zeta_* & -\lambda_* - (\zeta_* +\xi_*) k_*^2   \\
    -\lambda_*- k_*^2 {(\zeta_*+\xi_*)}  &  \lambda_*^2 + \lambda_* + ( \zeta_* + 2\xi_*+ Z) k_*^2  
\end{array}\right]=0 .
\end{equation} 

Equation \eqref{lambda_P_non_dim} defines a fourth-order polynomial for $\lambda_*$, thus, for a given $\bk_*$ there are 4 roots corresponding to the $P$-waves. Combining with 8 roots for $S$-waves, we get the total number of roots found being equal to 12, which is exactly the number of solutions for $\lambda_*(\bk_*)$ expected from \eqref{S_non_dim_simplified}. Thus, we have found all the roots of the equation \eqref{S_non_dim_simplified}. 
After computing the determinant in \eqref{lambda_P_non_dim} we get the following polynomial
\rem{ \begin{equation}
\begin{aligned} 
\label{poly_p} \delta \lambda_*^4   + \lambda_*^3(\delta + 1) &  + \lambda_*^2k_*^2 
\left(  \zeta_*+  \delta (Z +\xi_*+\zeta_* ) \right) \\
&+ \lambda_* k_*^2 (Z{-\xi_*}) + k_*^4 (\zeta_* Z {-\xi_*(\zeta_*+\xi_*)}) =0
\end{aligned} 
\end{equation}} 
\begin{equation}
\begin{aligned} 
\label{poly_p} 
\hspace{-1mm} \delta \lambda_*^4   \!+\!  \lambda_*^3(\delta + 1) &  \!+\!  \lambda_*^2k_*^2 
\left(  \zeta_* \!+\!  \delta (Z +2\xi_*+\zeta_* ) \right) 
+ \lambda_* k_*^2 Z \!+\!  k_*^4 (\zeta_* Z {-\xi_*^2}) \!=\!0 \, . 
\end{aligned} 
\end{equation}
For the stability of polynomial \eqref{poly_p} we investigate the principal minors $\Delta_i$, $i=1, \ldots 4$ of the Hurwitz matrix associated with the polynomial (Li\'enard-Chipart form of the criterion). 
The Hurwitz matrix corresponding to this polynomial has the form
\begin{equation} 
\left[ 
\begin{array}{cccc}  
\delta +1 & Z k_*^2 & 0 & 0 
\\ 
\delta & K_1 & K_2 & 0 
\\ 
0 & \delta+1 & Z k_*^2 & 0 
\\ 
0 & \delta & K_1 & K_2 
\end{array} 
\right] 
 \end{equation}
 where we have defined 
 \begin{equation} 
 \begin{aligned} 
 \label{K12_def} 
K_1= \left( \zeta_* (\delta+1)+\delta(Z+ 2 \xi_*)\right) k_*^2, \qquad 
K_2 = \left( \zeta_* Z - \xi_*^2 \right) k_*^4 \, . 
\end{aligned} 
\end{equation}
\rem{ 
As follows from the Routh-Hurwitz stability criterion, the polynomial  $s^4 + a_3 s^3 + a_2 s^2 + a_1 s +a_0$ is stable if  and only if $a_3 > 0, a_0 > 0$, 
$\Delta_2 := a_2a_3 - a_1 > 0,$ and $\Delta_3 := a_1\Delta_2 - a_0a_3^2 > 0$. Clearly, $1+\delta>0$ since $\delta>0$. 
Moreover, since $\mu_*>-1$ for the stability of $P$-waves, 
\begin{equation} 
Z=2 + \frac{\Lambda}{G} + 3 \mu_*>\frac{2(1-\nu)}{1-2 \nu}-3 > 2-3 >-1 \, . 
\label{Z_positive} 
\end{equation} But this result is not very useful for the stability analysis of the \eqref{poly_p}
} 
All $\Delta_i$ (with their exact forms given below) must be positive for stability. 
First, we notice that the conditions $\Delta_1>0$ and $\Delta_3>0$ read
\begin{equation}\Delta_1 = \delta + 1 > 0 \mbox{ and } \Delta_3 = (\delta Z + (\delta+1) \xi_*)^2 k_*^4 > 0,
\end{equation}
and are trivially satisfied. 

Next, we study the condition $\Delta_4>0$. Since $\Delta_3>0$, we can write, equivalently,    
\begin{equation} 
\frac{\Delta_4}{\Delta_3k_*^4} =     \zeta_*Z  -\xi_*^2> 0
\quad 
\Leftrightarrow 
\quad 
 \zeta_*\left(  \frac{\Lambda}{G}+2 + 3 \mu_* \right)  > \xi_*^2 \, . 
\label{Delta_4_stab_0} 
\end{equation} 
\rem{ 
we must have positive discriminant if consider dependence of $\xi$, i.e. \[\zeta_*^2 + 4Z\zeta_* > 0,\] or equivalently \[|\zeta + 2Z| > |2Z|,\] so, assuming $Z > 0,$ should be either $\zeta > 0,$ which is true, or $\zeta < -4Z,$ which is false, as can be proved to be inconsistent with $\Delta_2 > 0.$ In fact for the stability it is not necessary, that $Z>0,$ so if we assume $Z \le 0,$ we must have either $\zeta < 0,$ which is again inconsistent with $\Delta_2 > 0,$ or $\zeta > 4Z,$ and that is a possible case.
Next, using $\Delta_4 > 0,$ we express \[\xi = \frac{-\zeta + \alpha\sqrt{\zeta^2 + 4Z\zeta}}{2},\ \alpha \in [-1,1].\] W.l.o.g. $\zeta = 4\beta Z,\ \beta > 0,$ so 
$\xi = {2Z(-\beta + \alpha\sqrt{\beta^2 + \beta})}.$ We plug this into the expression for
} 
Finally, we compute the condition $\Delta_2>0$: 
\begin{equation} 
\label{Delta_2_cond} 
\begin{aligned} 
\frac{\Delta_2}{k_*^2} = \delta^2 Z &+ 2 \delta(\delta+1)\xi_* + (\delta+1)^2 \zeta_* .
\end{aligned} 
\end{equation} 
For stability of the steady state, we must have $\zeta_*>0$, otherwise $v=v_0$ is not a stable equilibrium. Multiplying condition \eqref{Delta_2_cond} 
by $\zeta_*$, and adding/subtracting the term $\delta^2 \xi_*^2$, we obtain an equivalent formulation 
\begin{equation} 
\frac{\Delta_2}{k_*^2}\zeta_*=\delta^2 \left( Z \zeta_*-\xi_*^2 \right) + \left( \delta \xi_*+ (\delta+1) \zeta_*\right)^2 >0 \, . 
\label{Delta_2_cond_mod}
\end{equation}
which is satisfied as long as \eqref{Delta_4_stab_0} is true. Since for physical reasons we necessarily have $G>0$, the stability condition for the $P$-waves can be rewritten as
\begin{equation} 
\zeta_*>0 \quad\text{and}\quad  
2 \left( G+ G \mu_*\right) 
+ 
\left( \Lambda +G  \mu_*- G \frac{\xi_*^2}{\zeta_*} \right) >  0 \, . 
\label{Delta_2_cond_mod_again}
\end{equation} 
Using the conditions \eqref{T_L_choice} and \eqref{zeta_star}, we can transform \eqref{Delta_2_cond_mod_again}  to the following form which will be useful for using the Sylvester criterion \eqref{Sylvester_porous} below: 
\begin{equation} 
2 \left( G+g_0 \mu\right) + \left( \Lambda + g_0\mu- g_0\frac{\xi^2}{\zeta} \right) >  0 \, . 
\label{Delta_2_cond_mod_yet_again}
\end{equation} 

We shall now show that the condition for the stability of the $P$-waves \eqref{Delta_4_stab_0} is exactly equivalent to the requirement for consistency of modified $P$-wave modulus in an isotropic medium. 

\paragraph{A digression: Linear stability of purely elastic media.} 
Let us now elucidate the physical meaning of \eqref{Delta_4_stab_0}, which, as we show, is simply the condition on the stability of propagation of $P$-waves in an elastic media. Suppose a wave is propagating in an elastic media with Lam\'e coefficients $(\Lambda,G)$ in accordance with \eqref{Hookeslaw}. 
The linearized equation for wave propagation is 
\begin{equation} 
\rho_s^0 \partial_t\delta\bu_s = \operatorname{div} \sigma_1(\epsilon)
\quad 
\Leftrightarrow 
\quad 
\lambda^2 \bv = - G \bv |\bk|^2 - (\Lambda + G)  (\bk \cdot \bv ) \bk \,,
\label{wave_consistency} 
\end{equation} 
where we assumed unstressed or relaxed elastic media \emph{i.e.} $ \sigma _0= \left.\pp{V}{b}\right|_0=0$ so that $ \mu=0$.

For $S$-waves, $\bv \perp \bk$, and $\lambda$ is purely imaginary if and only if $G>0$. For $P$-waves, $\lambda$ is purely imaginary if $2G + \Lambda>0$. The coefficient $2 G+ \Lambda$ is also known as the $P$-wave modulus of the elastic media. As we shall see, the condition of positive $P$-wave modulus will play the crucial part in the stability considerations.

For further discussion, it is interesting to compute the general condition on the convexity of the potential energy in the purely elastic case. In this case 
In this case \eqref{expansion_potential} reduces to
\begin{equation}\label{Pot_energy_elastic}
\begin{aligned}
V_0(b) &\simeq \frac{1}{2}(b-b_0) :\mathbb{C}: (b-b_0)
\simeq  \, G \epsilon:\epsilon + \frac12 \Lambda (\tr{\epsilon})^2 
\\
&=2 G \sum_{i>j} \epsilon_{ij}^2 
+
\frac12 \mathbf{X}_0^T \mathbb{Q}_0 
\mathbf{X}_0\, , \quad \mathbf{X}_0 := (\epsilon_{11}, \epsilon_{22}, \epsilon_{33}),
\end{aligned}  
\end{equation} 
and we have defined the quadratic form $\mathbb{Q}_0$ to be 
\begin{equation} 
\label{Q_elastic_def} 
\mathbb{Q}_0:= 
\left[
\begin{array}{ccc} 
2 G + \Lambda & \Lambda & \Lambda 
\\ 
\Lambda & 2 G + \Lambda & \Lambda 
\\ 
\Lambda & \Lambda & 2 G + \Lambda 
\end{array} 
\right].
\end{equation} 
Assuming that the coefficients $\epsilon_{ij}$ are independent numbers for a given deformations, the condition on $V_0$ to be positive definite is equivalent to the condition that the quadratic form $\mathbb{Q}_0$ is positive definite. By the Sylvester criterion, the quadratic form is positive definite if and only if all the leading principal minors are positive, leading to 
\begin{equation} 
\mbox{a) } \, 2 G + \Lambda>0 \, , \quad 
\mbox{b) } \, 2 G + 2 \Lambda >0 \, , \quad
\mbox{c) } \, 2 G + 3 \Lambda >0 \, . 
\label{Sylvester_elastic} 
\end{equation}
The first minor, \emph{i.e.}, condition a) is exactly the stability of $P$-waves. The third condition c) is equivalent to the positivity of the bulk modulus of the material. The second condition b) follows from the first and the third conditions. 

As we shall see immediately below, the conditions for the well-posedness of the $P$-waves and positive definite nature of the potential energy for the porous media follows closely the purely elastic framework, with the appropriate corrections due to the dynamics of the pores $v$. 

\paragraph{Justification of \eqref{Delta_4_stab_0} from the potential energy considerations.} 
Let us now consider the  case of a general potential energy $V(b,v)$ locally expressed about the equilibrium according to the quadratic expansion \eqref{expansion_potential}. Without loss of generality we consider the case with no linear terms, \emph{i.e.} $\mu = 0$, since the terms proportional to $\mu$ can be absorbed into $G$ and $\Lambda$ according to \eqref{Lame_shift}, with $G \rightarrow G+ \mu$ and $\Lambda \rightarrow \Lambda+\mu$. We have 
\begin{equation} 
\begin{aligned} 
V(b,v) &\simeq    \frac12 (b-b_0):\mathbb{C}:(b-b_0) + \frac{c_0 \zeta}{2v_0}(v-v_0)^2 + c_0 \xi \tr{\epsilon}(v-v_0)\\ 
&\simeq  G \epsilon:\epsilon + \frac12 \Lambda (\tr{\epsilon})^2 + \frac{c_0 \zeta}{2v_0}(v-v_0)^2 + c_0 \xi \tr{\epsilon}(v-v_0),
\end{aligned} 
\label{Pot_energy_rewritten} 
\end{equation}
hence $V$ is a quadratic form of $7$ variables: $(\epsilon_{ij})$ (6 elements from symmetry) and $(v-v_0)$. However, the off-diagonal elements of tensor $\epsilon$, namely $(\epsilon_{12},\epsilon_{23},\epsilon_{13})$ enter only in terms of squares multiplied by $G>0$. Thus, we rewrite 
\eqref{Pot_energy_rewritten} in the following form:  
\begin{equation} 
V(b,v) \simeq 2G\sum_{i>j}\epsilon^2_{ij} +\frac12 
\mathbf{X}^T \cdot  
\mathbb{Q}  \cdot \mathbf{X}, \quad 
\mathbf{X}:= \left(v-v_0,\epsilon_{11}, \epsilon_{22}, \epsilon_{33} \right)^T \, ,
\label{Pot_energy_rewritten_again} 
\end{equation}  
where we have defined a $4 \times 4$ quadratic form $\mathbb{Q}$ as 
\begin{equation} 
\label{Q_form_def} 
\mathbb{Q}:= \left[
\begin{array}{cccc}
\frac{c_0 \zeta}{v_0}  & c_0 \xi & c_0 \xi  & c_0 \xi 
\\ 
c_0 \xi & 2G+{\Lambda} & {\Lambda} & {\Lambda} \\
c_0 \xi & {\Lambda} & 2G+{\Lambda} & {\Lambda} \\
c_0 \xi & {\Lambda} & {\Lambda} & 2G+{\Lambda} 
\end{array}
\right].
\end{equation} 
Assuming the independence of all components of the strain tensor $\epsilon_{ij}$, we see that $V$ is a convex, positive definite function if and only if the quadratic form $\mathbb{Q}$ is positive definite. 

The Sylvester criterion gives four stability conditions: 
\begin{equation} 
\left\{ 
\begin{aligned} 
\Delta_1 & = \frac{c_0 \zeta}{v_0} > 0
\\ 
\Delta_2 & = \frac{c_0\zeta}{ v_0} 
\left( 2 G + \Lambda - \frac{g_0 \xi^2}{\zeta} \right) >0
\\ 
\Delta_3 & = \frac{4 G\zeta c_0}{ v_0 }
\left( G + \Lambda - 
\frac{g_0 \xi^2}{\zeta} \right) >0
\\ 
\Delta_4 &= 
\det \mathbb{Q} = \frac{4c_0G^2\zeta}{v_0}\left[ 2G + 3
\left(\Lambda- \frac{g_0 \xi^2}{\zeta} \right) \right] > 0\,,
\end{aligned} 
\right. 
\label{Sylvester_porous} 
\end{equation} 
where we recall that $g _0= c _0 v_0$.
The first condition of this system simply enforces the convexity of $V$ with respect to the small changes in $v$ about the equilibrium, and is thus very natural. To investigate the remaining three conditions, let us denote 
\begin{equation} 
\tilde{\Lambda}=\Lambda - \frac{g_0 \xi^2}{\zeta} \, . 
\label{tilde_Lam_def} 
\end{equation} 
We notice that the conditions for $\Delta_2>0$, $\Delta_3>0$ and $\Delta_4>0$
in \eqref{Sylvester_porous} are equivalent to the conditions 
\eqref{Sylvester_elastic} with the substitution $\Lambda \rightarrow \tilde{\Lambda}$.  Thus, the new variable defined by \eqref{tilde_Lam_def} acquires the physical meaning of the effective value of the second Lam\'e coefficient for the porous media. We remind the reader that the coefficients $\zeta$ and $\xi$ encode the values of the second derivatives of $V$ with respect to $v$ and $(v,b)$ respectively, and are thus appearing only in the description of the porous media. No corresponding values exist for the purely elastic media. It is thus even more surprising that the stability criteria for the porous media can be written in the form very similar to the elastic media through the combination of variable \eqref{tilde_Lam_def}. 

Note also that the condition for the $P$-wave stability \eqref{Delta_2_cond_mod_again} for a general $\mu_*$ can now be written using the shift \eqref{Lame_shift} as $\Delta_2>0$ in \eqref{Sylvester_porous}. The last condition of \eqref{Sylvester_porous}, \emph{i.e.}, $\Delta_4>0$, is equivalent to the requirement that the effective bulk modulus of a dry porous media is positive.

\section{Comparison with Biot's theory}

The dispersion relation $\mathbb{S}(\bu, \bv)^T$ described by \eqref{matr_def_isotropic} can be mapped to a system of linear PDEs. Let us assume, for simplicity, an isotropic media and take $\mathbb{K}=\beta \id$. We use the mapping of powers of $\bk$ to differential operators in Fourier space as $\bk \otimes \bk \rightarrow - \nabla {\rm div}$ and $|\bk|^2 \rightarrow - \Delta$ to get 
\begin{equation}\label{PDE_dispersion} 
\left\{ 
\begin{array}{l}
\displaystyle\vspace{0.2cm}\rho_f g_0 \frac{\partial^2}{\partial t^2} \bu +\beta \frac{\partial}{\partial t}(\bu - \bv) -g_0 \zeta \nabla {\rm div} \bu + g_0 (\zeta+ \xi)   \nabla {\rm div} \bv=\mathbf{0}\\
\displaystyle\vspace{0.2cm}\rho_s^0 \frac{\partial^2}{\partial t^2}\bv- \beta \frac{\partial}{\partial t}(\bu - \bv) + g_0 {(\zeta + \xi)} \nabla {\rm div} \bu\\
\displaystyle \qquad \qquad   - \left( g_0 (\zeta +2 \xi+  2\mu) + \Lambda+ G \right) \nabla {\rm div} \bv -\left( G + g_0 \mu \right) \Delta \bv = \mathbf{0} \,.
\end{array} 
\right. 
\end{equation} 
Note that the contribution from pressure in our system exactly cancel, which is reasonable, as the  pressure fluctuations generated by the motion of porous media in an internal force and thus must vanish. The corresponding Biot's system is given by 
\begin{equation} 
\label{linear_biot_system_0}
\hspace{-0.33cm}\left\{
\begin{array}{l}
\displaystyle\vspace{0.2cm}\frac{\partial^2}{\partial t^2}(\rho^{(f)}_{22} \bu+ \rho_{12} \bv) + \beta \frac{\partial}{\partial t}(\bu - \bv) - \nabla\div (R\bu+ Q\bv )  =  \mathbf{0}, \\
\displaystyle\vspace{0.2cm}\frac{\partial^2}{\partial t^2}(\rho^{(s)}_{11} \bv + \rho_{12} \bu) - \beta \frac{\partial}{\partial t}(\bu - \bv) - \nabla\div (Q\bu + P\bv) + N  \nabla \times \nabla \times  \bv = \mathbf{0}
\end{array} 
\right. 
\end{equation}
with $N$ being shear modulus of the skeleton and the fluid/elastic body, assumed to be the same. 
We shall note that Biot's equations is not directly applicable to an incompressible fluid, since the expressions for the variables $P,Q$ and $R$ in \eqref{linear_biot_system_0} involve explicitly the bulk modulus of the fluid.  However, if we proceed formally and use the equations from the literature and put $K_f= \infty$ for an incompressible fluid, the expressions for $P,Q$ and $R$ in terms of the bulk modulii of the porous skeleton $K_b$ and the elastic body itself $K_s$, see \textit{e.g.}, \cite{Fellah2004ultrasonic} are given by 
\begin{equation} 
\label{PQR_def} 
P=(1-g_0) K_s + \frac{4}{3} N \,, \quad Q= g_0 K_s \, , \quad R= \frac{g_0^2 K_s}{1- g_0 - K_b/K_s } \, . 
\end{equation} 
Let us turn our attention to our theory described in \eqref{PDE_dispersion}, where we have set $\rho_{12}=\rho_{21}=0$. The case of $\rho_{12} \neq 0$ an $\rho_{21} \neq 0$ can be easily incorporated by considering a more general inertia matrix in the Lagrangian.  There is also an exact correspondence between the friction terms. Thus, we need to compare the coefficients of the spatial derivative terms. A direct comparison between Biot's linearized system \eqref{linear_biot_system_0} and \eqref{PDE_dispersion} gives 
$R= g_0 \zeta$ by observing the coefficients of the terms proportional to $\nabla {\rm div} \bu$ from the equations \eqref{linear_biot_system_0}. From the term proportional to  $\nabla {\rm div} \bv$ in the first equation of \eqref{linear_biot_system_0}, we obtain $Q=-g_0 (\xi + \zeta)$. Finally by using $\nabla \times \nabla \times  \bv = \nabla \div \bv - \Delta \bv$ we obtain the expressions of $N$ and $P$. To summarize, the Biot's coefficients $(P,Q,R,N)$ are given by 
\begin{equation} 
\begin{aligned}
\label{Biot_correspondence}
R &= g_0\zeta,
\\
Q & = -g_0(\xi+\zeta),
\\
N & = G + g_0\mu,\   
\\
P&=(\Lambda+g_0 \mu )+2 (G+ g_0 \mu) + g_0 ( \zeta + 2 \xi) \,.
\end{aligned}
\end{equation}
Note that the expression $\Lambda+2 G$ is also known as the $P-$wave modulus. In our case, this $P$-wave modulus is modified by a shift of Lam\'e coefficients by $g_0 \mu$ and additional terms $\xi$ and $\zeta$  coming from the elasticity properties of the porous matrix. 

\rem{ 
\begin{equation} 
\label{Biot_coeff_eqs} 
\begin{aligned} 
Q&=-g_0 (\xi + \zeta) 
\\ 
R&=g_0 \xi 
\\ 
N &= g_0 \mu+\frac{E}{2(1+ \nu)} 
\\ 
P& =g_0 (\zeta+\xi + 2 \mu) + \frac{ (1-\nu) E}{(1+ \nu)(1-2 \nu)} \, .
\end{aligned} 
\end{equation} 
}

\section{Numerical investigation of phase and group velocities and attenuation}

We investigate the non-dimensionalized dispersion relations \eqref{s_wave_disp} and \eqref{poly_p} derived above in order to explore the phase velocity, group velocity, and attenuation coefficients of wave propagation in porous media. Instead of computing the roots $\lambda=\lambda(k)$, we consider the response of the system to a fixed frequency, as is common in the literature.  Thus, we take $\lambda=i \omega$ as a fixed parameter, and compute $k=k(\omega)$ from the dispersion relations. Then, the phase velocity is given by $v_p={\rm Re} \, \omega/k(\omega)$. Once $k(\omega)$ is known, we  compute the group velocity $v_g={\rm Re} \, d\omega/d k = {\rm Re} \,  (d k/d \omega)^{-1}$ by directly differentiating the dispersion relations as an implicit function and substituting $(\omega, k=k(\omega)) $. We also present the attenuation coefficient for the wave ${\rm Im} \, k(\omega)$ and attenuation per cycle ${\rm Im} \, k(\omega) /{\rm Re} \, k(\omega) $.  

\begin{figure}[htbp]
  \centering
 \includegraphics[width=0.8\textwidth]{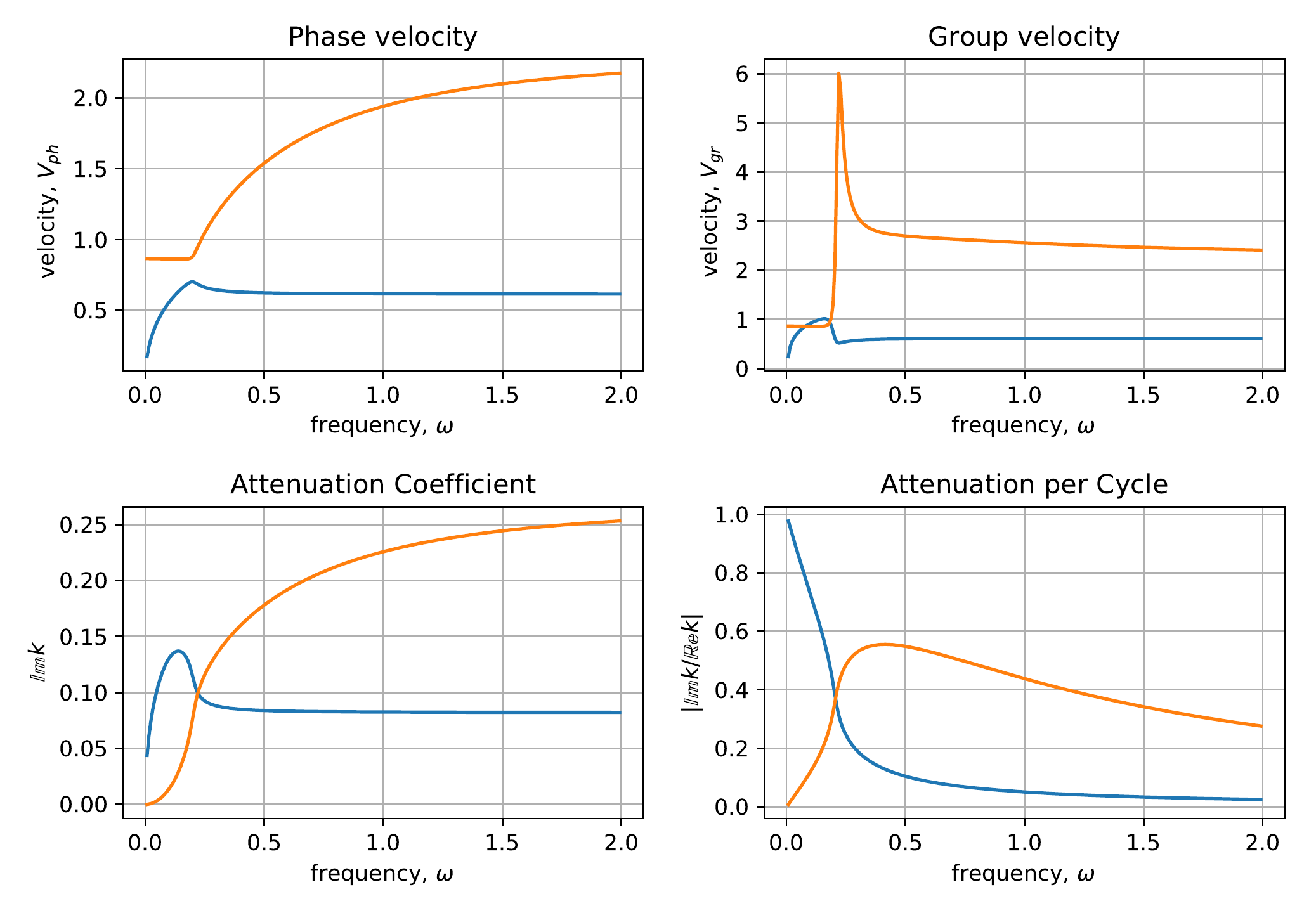}
  \caption{Velocities and attenuation coefficients for  $P$-waves  with $\delta = 3$, $Z=3$, $\zeta_*=2$ and $\xi_*=0$.  
  \label{fig:p0} }
\end{figure}

\begin{figure}[htbp]
  \centering
\includegraphics[width=0.8\textwidth]{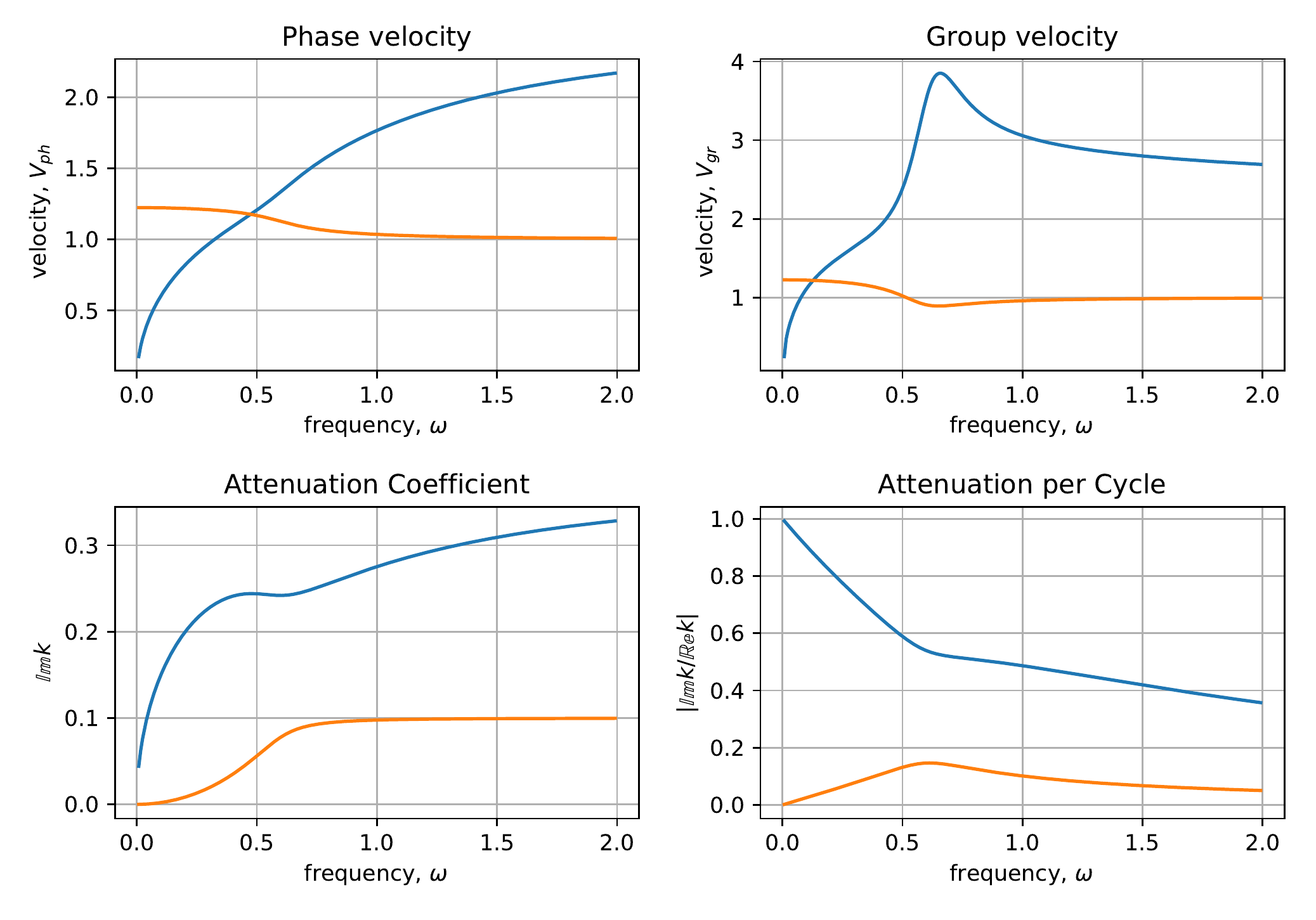}
  \caption{Velocities and attenuation coefficients for  $P$-waves with $\delta = 1$, $Z=3$, $\zeta_*=2$ and $\xi_*=0$. 
    \label{fig:p2} }
\end{figure}

\begin{figure}[htbp]
  \centering
\includegraphics[width=0.8\textwidth]{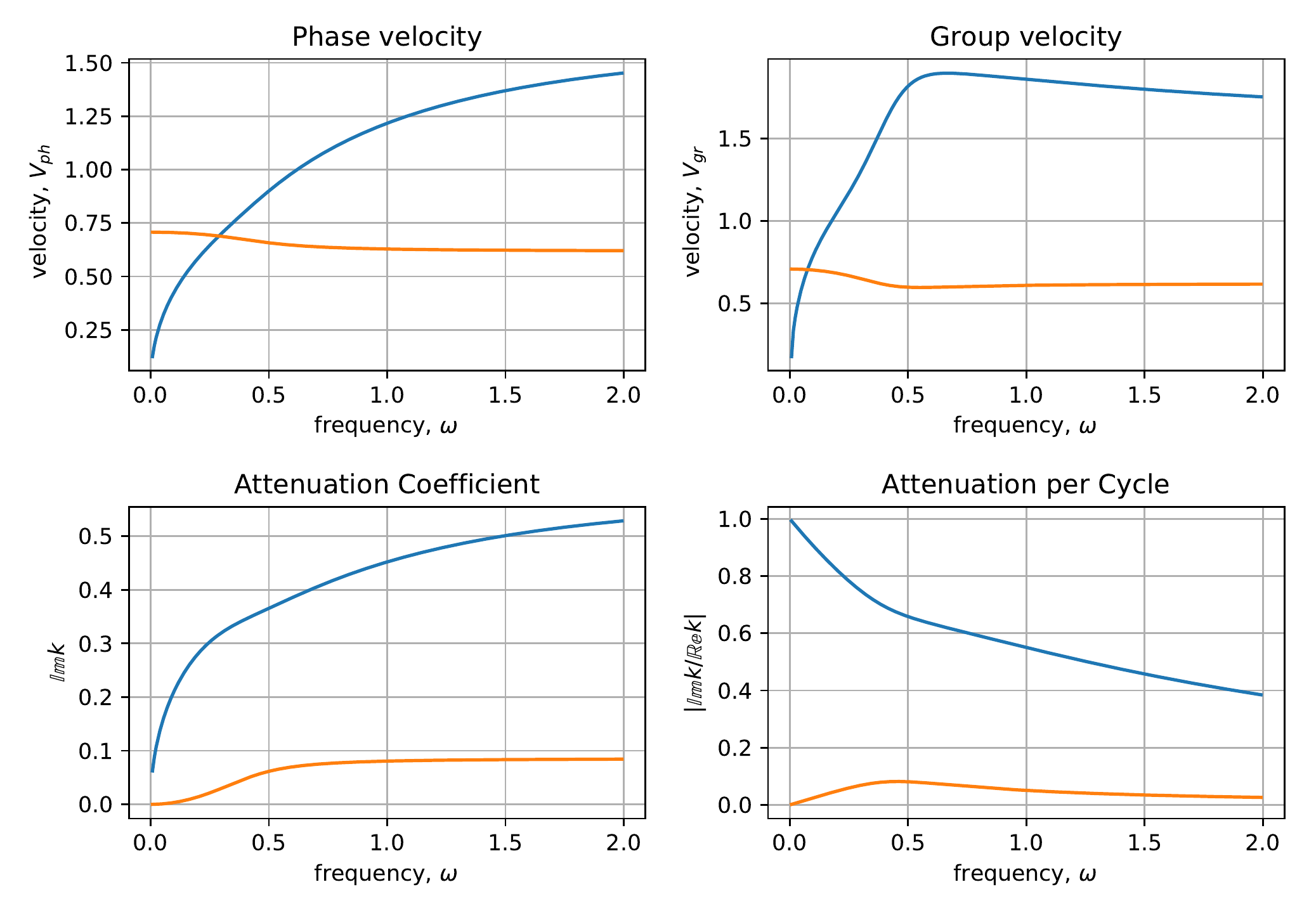}
  \caption{Velocities and attenuation coefficients for  $P$-waves with $\delta = 1$, $Z=1$, $\zeta_*=1$ and $\xi_*=1$.  
    \label{fig:p4}}
\end{figure}

\begin{figure}[htbp]
  \centering
 \includegraphics[width=0.8\textwidth]{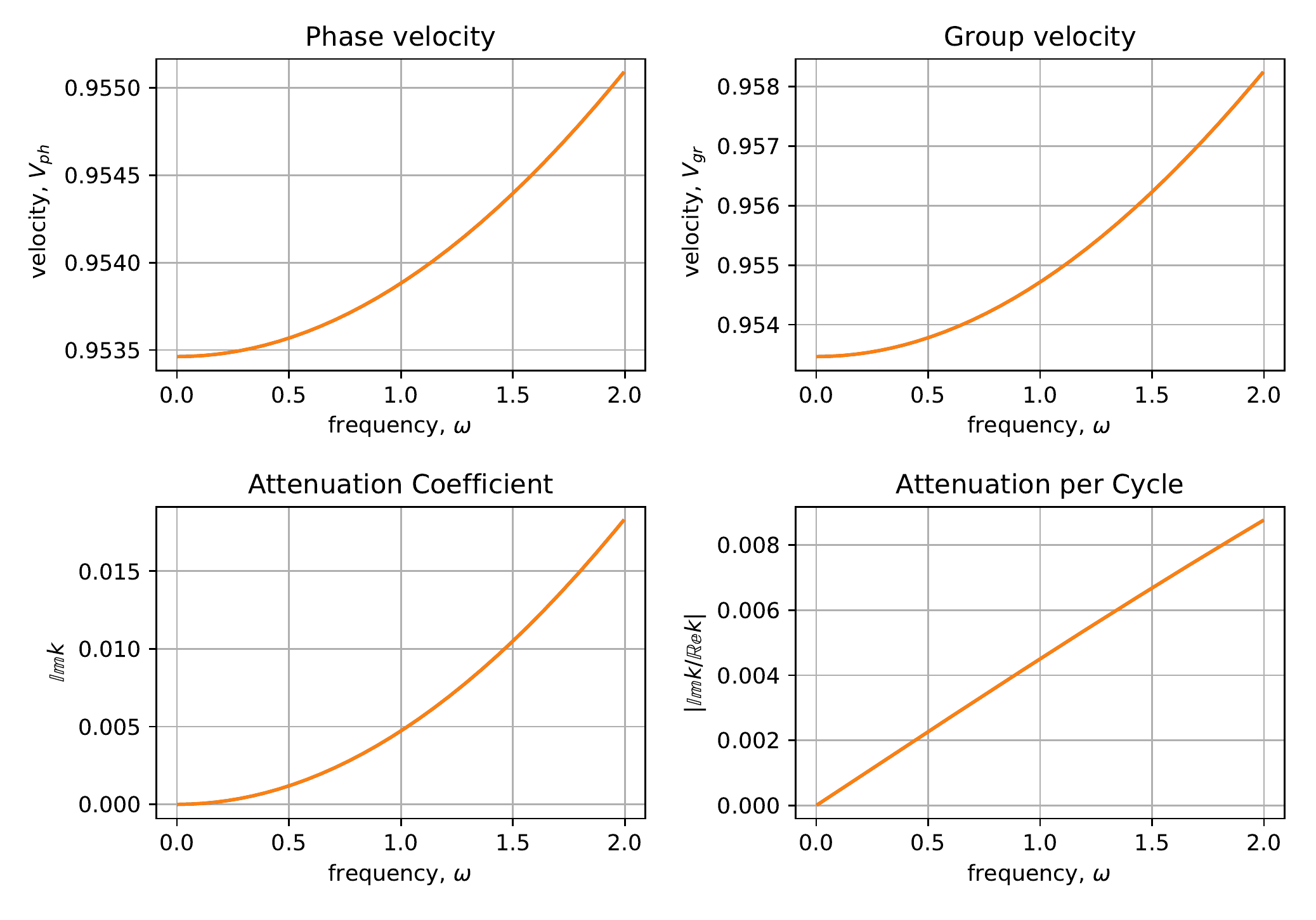}
  \caption{Velocities and attenuation coefficients for  $S$-waves  with $\delta = 0.1$ and $\mu_*=0$.  
    \label{fig:s0}}
\end{figure}

\begin{figure}[htbp]
  \centering
 \includegraphics[width=0.8\textwidth]{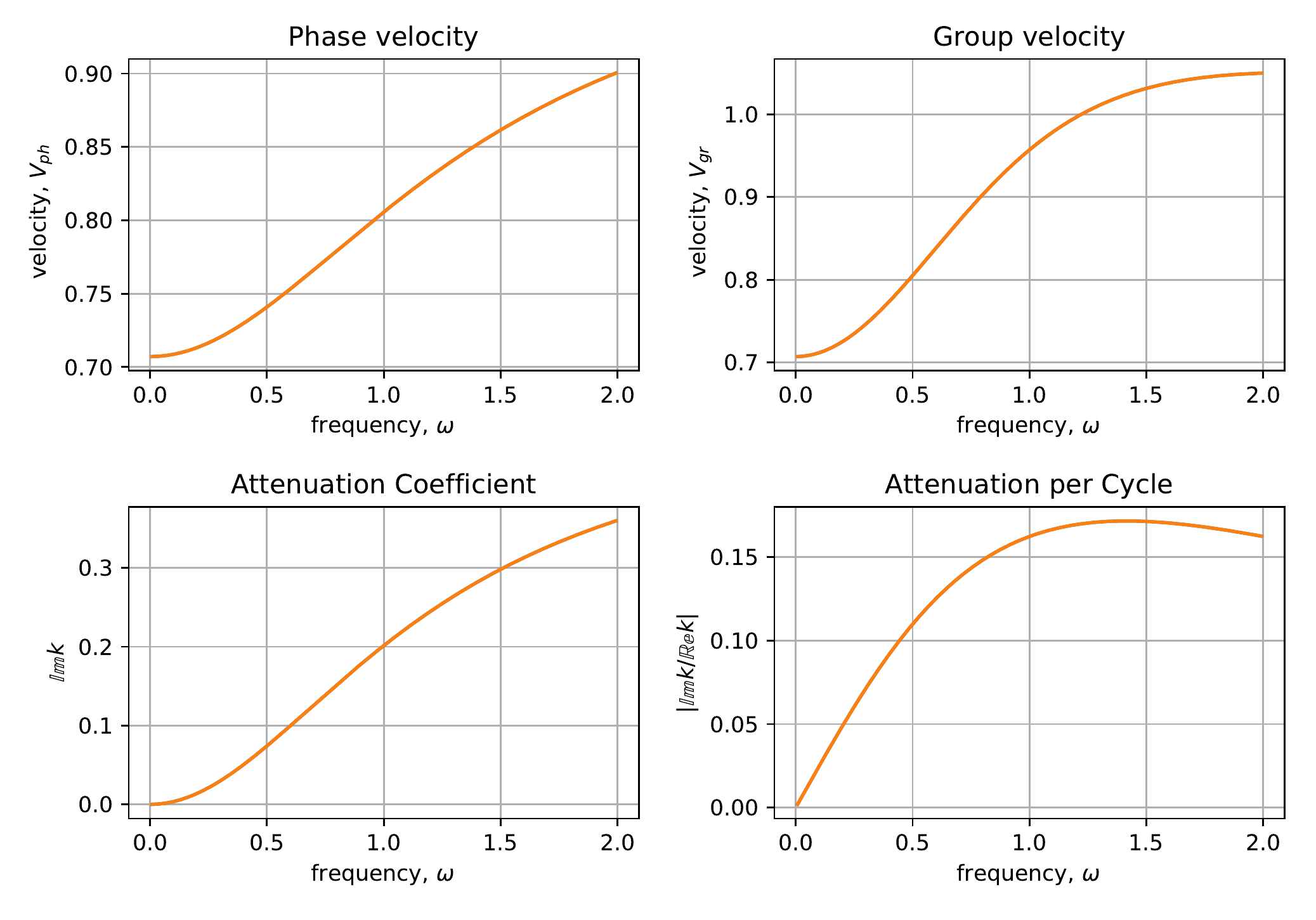}
  \caption{Velocities and attenuation coefficients for  $S$-waves  with $\delta = 1$ and $\mu_*=0$. 
     \label{fig:s2} }
\end{figure}

\begin{figure}[htbp]
  \centering
 \includegraphics[width=0.8\textwidth]{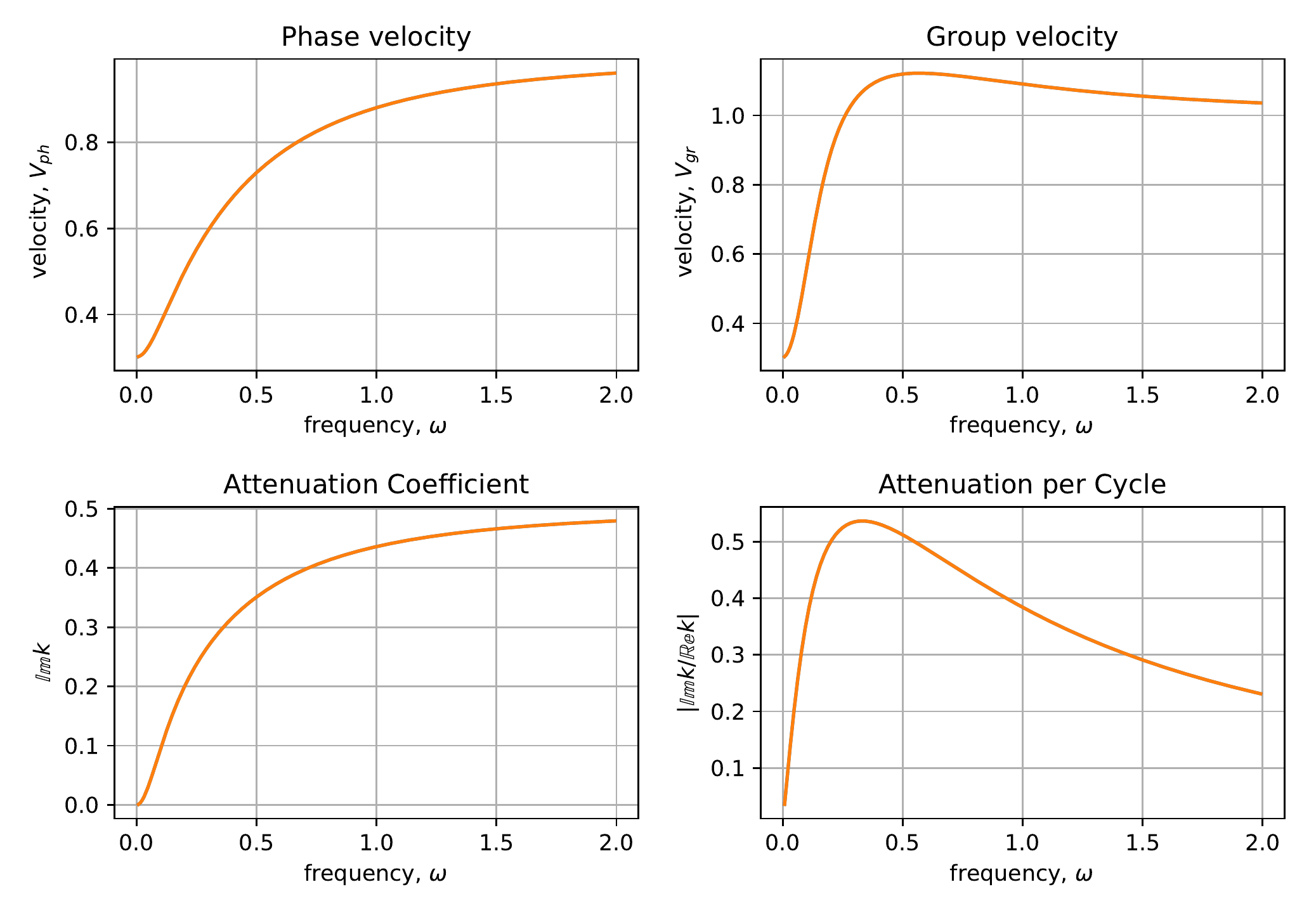}
  \caption{Velocities and attenuation coefficients for  $S$-waves  with $\delta = 10$ and $\mu_*=0$. 
     \label{fig:s4} }
\end{figure}

According to \eqref{zeta_star}, and the fact that $\zeta$ has the order of magnitude of the microscopic bulk modulus, most materials will have $\zeta_* \sim 1$, $\xi_* \sim 1$, $Z \sim 1$ at the order of magnitude.  For biological materials, $\delta$ tends to be large whereas for  porous media made out of dense materials conveying gas, $\delta$ is small. We thus explore both large and small values of $\delta$ in the simulations. In Figures~\ref{fig:p0}-\ref{fig:p4} we present the results of computation of dispersion relation for the $P$-waves for a set of different parameters $\delta$, $Z$, $\zeta_*$ and $\xi_*$. Only two roots of equation \eqref{poly_p} are shown since the equation is a quadratic equation in $k^2$. The other roots correspond to the waves propagating with the same velocity and attenuation coefficient in the opposite direction.  In Figures~\ref{fig:s0}-\ref{fig:s4} we present the results of computation of dispersion relation for the $S$-waves for a set of different parameter $\delta$. The axes variables in the figures are dimensionless, rescaled according to the time and length scales defined in \eqref{T_L_choice}. 

\section{Conclusions}

In this paper, we have derived the equations of motion for a porous media filled with an incompressible fluid. We have chosen to write all the equations in the Eulerian frame for both the fluid and the porous media. Our equations are valid for arbitrary deformations, and, as far as we are aware, are new. We have compared the linearized equations of motion to the Biot's equations and found a correspondence between our equations and Biot's equations, with a clear and physical interpretation of the parameters. We have also derived the stability of the linearized system for the porous system which turned out to be equivalent to the requirements that the parameters of the dry porous media being physically consistent. 

For further studies, it will be interesting to consider the dynamics of active porous materials like sea sponges. Recent work \cite{ludeman2014evolutionary} has demonstrated interesting 'sneezing' dynamics of a freshwater sponge, when the sponge contracts and expands to clear itself from surrounding polluted water. Equations \eqref{expressions_explicit} can be readily modified to model such contraction, for example, by making the equilibrium value of $v=v_0$ a prescribed function of time. We believe that a semi-analytic theory may be developed in that case under the assumption of radial or spherical symmetry, which is sufficiently close to the experimental case. This problem, as well as other interesting topics in active, fluid-filled porous media, will be considered in our upcoming work. 

\paragraph{acknowledgements.} 
We are thankful for fruitful and productive discussions with Profs G. L. Brovko, D. D. Holm, A. Ibraguimov, T. S. Ratiu and D. V. Zenkov. VP is also thankful for the lively and informative discussion with the participants of G. L. Brovko's seminar on Elasticity Theory at the Moscow State University (November 2018), where a preliminary version of this work was presented. FGB is partially supported by the ANR project GEOMFLUID 14-CE23-0002-01.  TF and VP were partially supported by NSERC and University of Alberta.

\end{document}